\documentclass[3p,times,11pt]{elsarticle}
\usepackage{setspace}
\usepackage{lipsum}
\usepackage{mdsymbol}
\usepackage{amsthm}
\usepackage{caption,subcaption}
\usepackage{dutchcal}
\usepackage{threeparttable}
\usepackage{array}
\usepackage{multirow}
\usepackage{algorithmic}
\usepackage{algorithm}
\makeatletter
\def\ps@pprintTitle{%
	\let\@oddhead\@empty
	\let\@evenhead\@empty
	\def\@oddfoot{}%
	\let\@evenfoot\@oddfoot}
\makeatother
\usepackage{lineno,hyperref,booktabs}
\modulolinenumbers[5]

\journal{arXiv preprint}

\begin{document}

\begin{frontmatter}

\title{Complex Isotropic $\alpha$-Stable-Rician Model\\ for Heterogeneous SAR Images}

\author{Mutong LI}

\author{Ercan Engin Kuruoglu}

\address{Tsinghua-Berkeley Shenzhen Institute, Tsinghua University}

\begin{abstract}
This article introduces a novel probability distribution model, namely Complex Isotropic $\alpha$-Stable-Rician (CI$\alpha$SR), for characterizing the data histogram of synthetic aperture radar (SAR) images. Having its foundation situated on the Lévy $\alpha$-stable distribution suggested by a generalized Central Limit Theorem, the model promises great potential in accurately capturing SAR image features of extreme heterogeneity. A novel parameter estimation method based on the generalization of method of moments to expectations of Bessel functions is devised to resolve the model in a relatively compact and computationally efficient manner. 
Experimental results based on both synthetic and empirical SAR data exhibit the CI$\alpha$SR model's superior capacity in modelling scenes of a wide range of heterogeneity when compared to other state-of-the-art models as quantified by various performance metrics. Additional experiments are conducted utilizing large-swath SAR images which encompass mixtures of several scenes to help interpret the CI$\alpha$SR model parameters, and to demonstrate the model's potential application in classification and target detection. 

\end{abstract}


\end{frontmatter}

\section{Introduction}
Statistical modelling of SAR image data has attracted unanimous attention and effort since the very introduction of this potent radar technology. A successful model does not only contribute to a faithful representation of the image data population, but also lays foundation for post-processing applications such as terrain classification \cite{tison2004new} and auto target detection \cite{copsey2003bayesian,gao2016scheme}. Amongst SAR images of various terrains, heterogeneous scenes such as urban areas and sea surface with ships are the most problematic for models to tackle, since steel plates and concrete constructions prove to be an exceptionally dominant reflector of radio waves. These images generally contain a good proportion of pixels with extremely high values, contributing to a long, thick tail in the data histogram.

SAR models with assorted motivations have been designed alongside the development of SAR technology itself.
The empirical models such as Weibull distribution \cite{ulaby1986weibull} and log-normal distribution \cite{george1968lognormal} are famed for their ability to capture SAR image data characteristics in rather homogeneous scenes.
The $\mathcal{K}$ family \cite{jakeman1976model} and its special class $\mathcal{G}^{0}$ \cite{frery1997model} are derived as the product model of both texture and speckle distributions for more heterogeneous images. 
In the context of this paper, we would like to emphasize on complex isotropic models induced by the Rayleigh distribution. The Rayleigh distribution \cite{strutt1894theory} has been the classical model for problems involving complex wave reflections in applications including telecommunications, sensing and radar imaging. Making the assumption of Gaussian real and imaginary components as suggested by the Central Limit Theorem, which states the distribution of a collection of infinitesimal reflectors with finite variance , the amplitude is obtained as Rayleigh which forms the simplest complex isotropic model. 
Yet, Rayleigh model assumes zero-mean reflections which is violated in the case of a dominant reflector in the scene. To accommodate said scenario, the model is generalized to the Rician model \cite{rice1944mathematical,nicolas2019new}, which has an additional location parameter that reflects the existence of dominant reflectors. By considering the data population as an amplitude distribution of a complex isotropic random variable, these models practically mimic the amplitude and phase constitution of complex radar signal.

While the Rician model takes care of a small or a single dominant reflector in an otherwise homogeneous scene, 
in the case of heterogeneous SAR images such as urban scenes, reflectance to radio waves by certain surface types such as automobiles or buildings becomes so strong that these reflectors’ contribution to the entire population can no longer be considered as infinitesimal. As a result, the Rician model established upon said assumptions collapses.

A good number of models have been proposed to intentionally cater for the heavy-tailed histogram by replacing the Gaussian distribution in the Rician model with a generalized version. One of the options for substitution is the Generalized Gaussian (GG), aka exponential power distribution. By introducing an additional shape parameter $\alpha$ to exponential argument of conventional Gaussian probability density function (pdf), the GG distribution is capable of accommodating histogram tails of varying thickness. GG distribution includes Gaussian and Laplace distribution as its special cases, its pdf is given as follows.
\begin{equation}
    f\left( x \middle| \alpha,\gamma,\delta \right) = \frac{\alpha}{2\gamma\Gamma\left( \frac{1}{\alpha} \right)}{\exp\left( {- \left| \frac{x - \delta}{\gamma} \right|^{\alpha}} \right)}
\end{equation}
Moser et al. \cite{moser2006sar} first demonstrated a SAR model named Generalized Gaussian-Rayleigh as the amplitude distribution of independent real and imaginary components with zero-mean GG random variables. The GG-Rayleigh model was later generalized by Karakus et al. \cite{karakucs2021generalized} into Generalized Gaussian-Rician (GGR) by including the missing location parameter $\delta$. The GGR model fully uncovers the potential of GG distribution, and includes other competetive models such as Nakagami-Rice \cite{dana1986impact} and Laplace-Rician \cite{karakucs2020LaplaceRician} as its special case. However, despite GGR's capability of modelling SAR images of a considerably wide range of heterogeneity, the model still fails to accommodate extremely heterogeneous urban data, and a lack of analytical parameter estimator also restricted the implementation of the model. The probability density function of GGR can be expressed as follows:
\begin{equation}
    \begin{split}
        f\left( x \middle| \alpha,\gamma,\delta \right) = 
        \frac{\alpha^{2}x}{4\gamma^{2}\Gamma^{2}\left( \frac{1}{\alpha} \right)}{\int_{0}^{2\pi}{{\exp\left( {- \frac{\left| {x{\cos\theta} - \delta} \right|^{\alpha} + \left| {x{\sin\theta} - \delta} \right|^{\alpha}}{\gamma^{\alpha}}} \right)}~d\theta}}
    \end{split}
\end{equation}

Another strain of complex isotropic model family is established upon Paul Lévy’s $\alpha$-stable distribution \cite{mandelbrot1960pareto,zolotarev1986one}. The $\alpha$-stable distribution $
S_{\alpha}\left( {\gamma,~\beta,~\delta} \right)$ adopts 4 parameters $\alpha$, $\beta$, $\gamma$, $\delta$ to characterize its impulsiveness, skewness, scale, and location, respectively, and is commonly presented with its characteristic function (CF) due to a lack of compact analytical pdf. The $\alpha$-stable distribution includes Gaussian ($\alpha$=2), Cauchy ($\alpha$=1, $\beta$=0), and Levy distribution ($\alpha$=0.5, $\beta$=1) as its special cases, the only ones with a compact analytical pdf. The CF of $\alpha$-stable distribution is given as follows:
\begin{equation}
    \begin{split}
        {\varphi}\left( \omega \middle| \alpha,\beta,\gamma,\delta \right) = 
        \left\{ \begin{matrix}
{{\exp\left\{ {j\delta\omega - \gamma^{\alpha}|\omega|^{\alpha}\left\lbrack {1 + j\beta sgn(\omega){\tan\left( \frac{\alpha\pi}{2} \right)}} \right\rbrack} \right\}},~\textup{if}~\alpha \neq 1} \\
{{\exp\left\{ {j\delta\omega - \gamma|\omega|\left\lbrack {1 + j\beta sgn(\omega)\frac{2}{\pi}{\log|\omega|}} \right\rbrack} \right\}},~\textup{if}~\alpha = 1} \\
\end{matrix} \right.
    \end{split}
\end{equation}
One immediate advantage of adopting $\alpha$-stable distribution in a complex isotropic SAR model lies in the distribution's capacity of covering Gnedenko and Kolmogorov's generalization of the the Central Limit Theorem (GCLT), which states that summation of large numbers of random variables of a distribution with power-shaped Paretian tails will tend to $\alpha$-stable distribution \cite{gnedenko&kolmogorov1968}. The incorporation of a heavier tail in $\alpha$-stable coincides with the impulsive feature of SAR images of heterogeneous scenes, therefore, making the distribution a promising choice to model the real and imaginary parts of a complex radar signal.

Kuruoglu and Zerubia \cite{kuruoglu2004modeling} first devised a model named Heavy-Tailed Rayleigh (HTR) which is the amplitude distribution of a complex isotropic $\alpha$-stable random variable with arbitrary characteristic exponent $\alpha$ and scale parameter $\gamma$, the model proved successful in modelling the tail of urban SAR data, and was also adjusted by Achim et al. \cite{achim2006sar} to filter speckle. The pdf of HTR is shown as follows.
\begin{equation}
    f\left( x \middle| \alpha,\gamma \right) = x{\int_{0}^{\infty}{{{\omega~\exp}\left( - \gamma^{\alpha}\omega^{\alpha} \right)}J_{0}(\omega x)~d\omega}}
\end{equation}
Later, Karakus et al. \cite{karakucs2022cauchy} developed the Cauchy-Rician model (CR) which utilizes Cauchy (special case of $\alpha$-stable with arbitrary $\gamma$ and $\delta$ parameter) as a designated treatment to the non-zero mean reflection in extremely heterogeneous urban scene. CR model made amends for HTR’s incapability in modelling urban areas with dominant reflectors such as large residential blocks, but at the cost of losing the characteristic exponent $\alpha$, an important degree of freedom giving flexibility in modelling various levels of impulsiveness.
Cauchy-Rician is frequently more impulsive than the actual SAR data. The pdf of CR model is shown as follows:
\begin{equation}
    \begin{split}
        f\left( x \middle| \gamma,\delta \right) =
        \frac{\gamma x}{2\pi}{\int_{0}^{2\pi}\frac{d\theta}{\left\lbrack {\gamma^{2} + x^{2} + 2\delta^{2} + 2x\delta\left( {\cos\theta} + {\sin\theta} \right)} \right\rbrack^{3/2}}}
    \end{split}
\end{equation}

In this article, we would like to propose a generalization to the HTR model, producing a novel SAR model named the Complex Isotropic $\alpha$-Stable Rician (CI$\alpha$SR) which includes the characteristic exponent, scale parameter, and location parameter from the $\alpha$-stable family. It combines the advantages of varying tail thickness in the HTR model and characteristic of non-zero-mean reflections in CR and Rician models, and manifests great potential in accurately modelling heterogeneous SAR images. Details regarding the model is arranged in the following manner: Section II provides a mathematical derivation of the model; Section III introduces a quasi-analytical parameter estimator for the model based on 
a generalized method of moments; Section IV exemplifies practical significance of the model by applying it to SAR images of various heterogeneity; Section V concludes the paper and discusses potential future work.

\section{Complex Isotropic $\alpha$-Stable Rician Model}
The CI$\alpha$SR model proposed in this work models the amplitude distribution of a complex isotropic $\alpha$-stable random variable. The derivation starts by generalizing complex symmetric $\alpha$-stable distribution \cite{cambanis1982complex} with an additional location parameter $\delta$. The characteristic function of an isotropic bivariate $\alpha$-stable distribution with characteristic exponent $\alpha$, scale parameter $\gamma$, and location parameter $\delta$ is
\begin{equation}
	\varphi\left( {\omega_{1},~\omega_{2}} \right) = \exp\left\lbrack j\left( \delta_{1}\omega_{1} + \delta_{2}\omega_{2} \right) - \gamma\left| \vec{\omega} \right|^{\alpha} \right\rbrack,
\end{equation}
where $\vec{\omega} = \omega_{1} + j\omega_{2}$ is the bivariate vector in frequency domain. $\delta = \sqrt{\delta_{1}^{2} + \delta_{2}^{2}}$ is the location parameter, with $\delta_{1}$ and $\delta_{2}$ arbitrarily allocated to the real and imaginary part. By performing the inverse Fourier transform of the above characteristic function, one acquires the pdf of the real and imaginary parts of the signal amplitude.
\begin{equation}
	\begin{split}
		{f}_{X_{re},~X_{im}}\left( {x_{re},~x_{im}} \right) =\frac{1}{4\pi^{2}}{\iint\limits_{\omega_{1},~\omega_{2}}^{~}{{\exp\left\lbrack {j\left( {\delta_{1}\omega_{1} + \delta_{2}\omega_{2}} \right) - \gamma^{\alpha}\left| \vec{\omega} \right|^{\alpha}} \right\rbrack}{\exp\left\lbrack {j\left( {\omega_{1}x_{re} + \omega_{2}x_{im}} \right)} \right\rbrack}d\omega_{1}d\omega_{2}}}
	\end{split}
\end{equation}
By assuming $\left. \omega = \middle| \vec{\omega} \middle| = \sqrt{\omega_{1}^{2} + \omega_{2}^{2}} \right.$ and $\theta_{\omega} = {\arctan\frac{\omega_{2}}{\omega_{1}}}$, the above integral can be converted into the polar coordinates.
\begin{equation}
	\begin{split}
		&{f}_{X_{re},~X_{im}}\left( {x_{re},~x_{im}} \right) = \\
		&\frac{1}{4\pi^{2}}{\int_{0}^{\infty}{{{{\omega}\exp}\left( - \gamma^{\alpha}\omega^{\alpha} \right)}{\int_{0}^{2\pi}{\exp\left\lbrack j\left( {\delta_{1}\omega_{1} + \delta_{2}\omega_{2}} \right) \right\rbrack}}}}\times {\exp\left\lbrack j\left( {\omega_{1}x_{re} + \omega_{2}x_{im}} \right) \right\rbrack}d\theta_{\omega}~d\omega
	\end{split}
\end{equation}
By further assuming $x = \sqrt{x_{re}^{2} + x_{im}^{2}}$ and $\theta_{x} = {\arctan\frac{x_{im}}{x_{re}}}$, one can convert the pdf from that of real and imaginary parts to amplitude and phase.
\begin{equation}
	\begin{split}
		{f}_{X,~\Theta_{x}}\left( {x,~\theta_{x}} \right) =& x {p}_{X_{re},~X_{im}}\left( {x{\cos\theta_{x}},~x{\sin\theta_{x}}} \right) \\
		= &\frac{x}{4\pi^{2}}{\int_{0}^{\infty}{{\omega~\exp}\left( {- \gamma^{\alpha}\omega^{\alpha}} \right)}}{\int_{0}^{2\pi}{\exp\left\lbrack {j\omega\left( {\delta_{1}{\cos\theta_{\omega}} + \delta_{2}{\sin\theta_{\omega}}} \right)} \right\rbrack}}\\
		&\times {\exp\left\lbrack {j\omega x\left( {{\cos\theta_{x}}{\cos\theta_{\omega}} + {\sin\theta_{x}}{\sin\theta_{\omega}}} \right)} \right\rbrack}d\theta_{\omega}~d\omega
	\end{split}
\end{equation}
Utilize the identity ${\cos A} \ {\cos B} + {\sin A} \ {\sin B} = {\cos(A - B)}$, and integrate the pdf given above with respect to $\Theta_{x}$ to obtain the marginal pdf of signal amplitude $X$.
\begin{equation}
	\begin{split}
		{f}_{X}(x) =& {\int\limits_{\theta_{x}}^{~}{{{\rm pdf}}_{X,~\Theta_{x}}\left( {x,~\theta_{x}} \right)~d\theta_{x}}}\\
		=& \frac{x}{4\pi^{2}}{\int_{0}^{\infty}{{\omega~\exp}\left( {- \gamma^{\alpha}\omega^{\alpha}} \right)}}{\int_{0}^{2\pi}{{\exp\left\lbrack {j\omega\left( {\delta_{1}{\cos\theta_{\omega}} + \delta_{2}{\sin\theta_{\omega}}} \right)} \right\rbrack}~}}\\
		&{\int_{0}^{2\pi}{{\exp\left\lbrack {j\omega x{\cos\left( {\theta_{x} - \theta_{\omega}} \right)}} \right\rbrack}~d\theta_{x}~d\theta_{\omega}~d\omega}}
	\end{split}
\end{equation}
By invoking the identity ${{a{\cos}}x} + b{\sin x} = \sqrt{a^{2} + b^{2}}{{~\sin}\left( x + {\arctan\frac{a}{b}} \right)}$ to merge the trigonometric term within the exponential term, and zeroth-order Bessel function of the first kind $
J_{0}(x) = \frac{1}{2\pi}{\int_{0}^{2\pi}{{\exp\left( {jx{\sin\theta}} \right)}~d\theta}}$, the above pdf can be reduced into a relatively compact form:
\begin{equation}
	{f}_{X}(x) = x{\int_{0}^{\infty}{{{\omega~\exp}\left( - \gamma^{\alpha}\omega^{\alpha} \right)}~J_{0}(\omega\delta)~J_{0}(\omega x)~d\omega}}.
	\label{eq.CIαSRpdf}
\end{equation}
This integral-form equation \ref{eq.CIαSRpdf} gives us the expression for the CI$\alpha$SR pdf. A careful reader may readily notice its resemblance with its degenerate version known as Heavy-Tailed Rayleigh which is obtained for $\delta=0$. As a simple check, the model can be easily reduced to Rician distribution at $\alpha$=2 with the following identity \cite{gradshteyn2014table}, just as a univariate $\alpha$-stable distribution degenerates to Gaussian.
\begin{equation}
	\begin{split}
		{\int_{0}^{\infty}{x{\exp\left( - \rho^{2}x^{2} \right)}J_{p}\left( {\alpha x} \right)J_{p}\left( {\beta x} \right)~dx}} = \frac{1}{2\rho^{2}}{\exp\left( - \frac{\alpha^{2} + \beta^{2}}{4\rho^{2}} \right)}I_{p}\left( \frac{\alpha\beta}{2\rho^{2}} \right)\\
		\left\lbrack {\rm{Re}}~p > - 1,\left| {\arg\rho} \right| < {\pi/4},~\alpha > 0,~\beta > 0 \right\rbrack
	\end{split}
\end{equation}

\begin{figure}
	\centering
	\subcaptionbox{\label{Fig_CIαSRModelCurve_sub1}}
	{\includegraphics[width=0.32\linewidth]{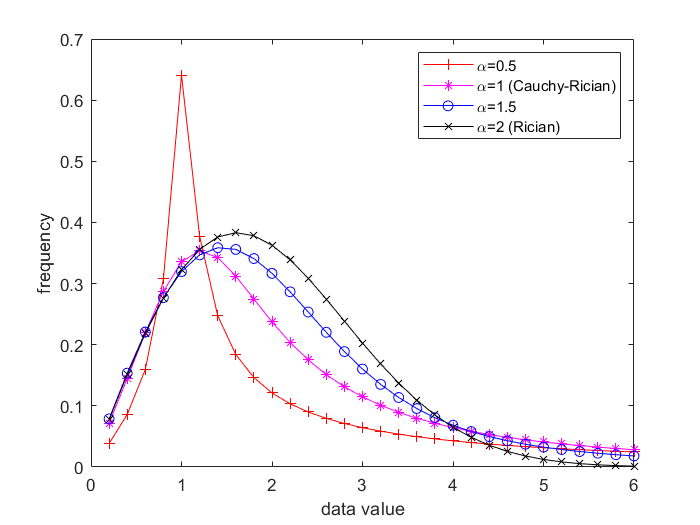}}
	\subcaptionbox{\label{Fig_CIαSRModelCurve_sub2}}
	{\includegraphics[width=0.32\linewidth]{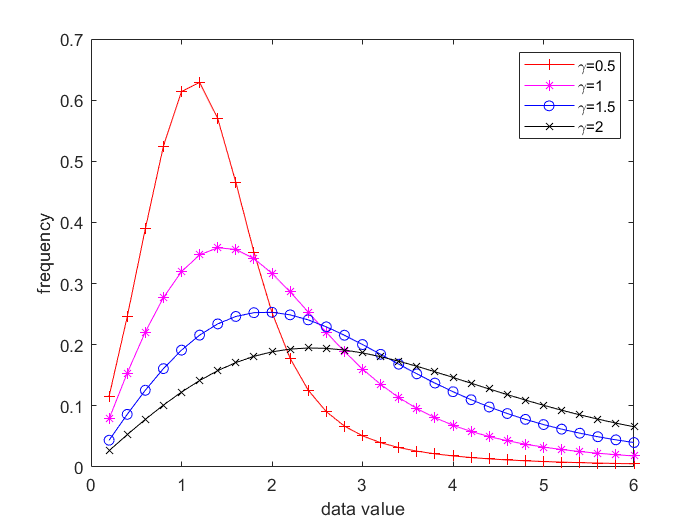}}
	\subcaptionbox{\label{Fig_CIαSRModelCurve_sub3}}
	{\includegraphics[width=0.32\linewidth]{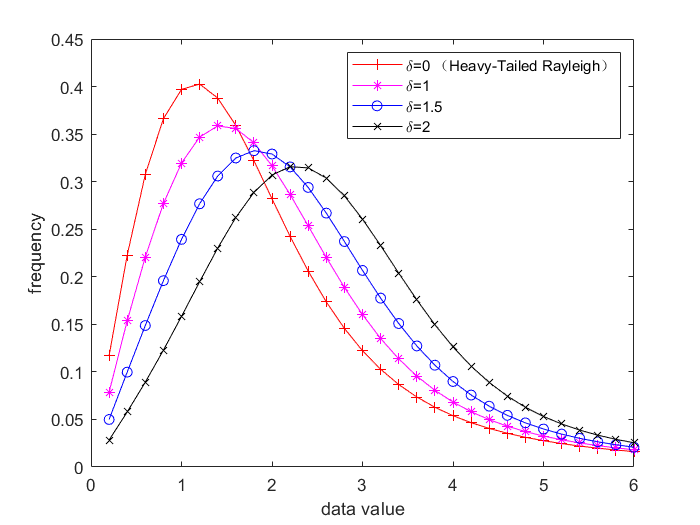}}
	\caption{CI$\alpha$SR model curves with varying parameter of (a) characteristic exponent $\alpha$ ($\gamma$=1, $\delta$=1); (b) scale parameter $\gamma$ ($\alpha$=1.5, $\delta$=1); and (c) location parameter $\delta$ ($\alpha$=1.5, $\gamma$=1).}
	\label{Fig_CIαSRModelCurve}
\end{figure}

Figure \ref{Fig_CIαSRModelCurve} are placed hereby to offer an outright example of how the model pdf varies according to changes of each parameter. The figure also provides an insight into how CI$\alpha$SR model includes Heavy-Tailed Rayleigh, Cauchy-Rician, and Rician as its special cases.
Figure \ref{Fig_CIαSRModelCurve_sub1} shows that characteristic exponent $\alpha$ decides the heaviness of the tail, while Figure \ref{Fig_CIαSRModelCurve_sub2} shows that scale parameter $\gamma$ sees how dispersed the peak is when spreading on the x-axis, and Figure \ref{Fig_CIαSRModelCurve_sub3} shows that location parameter $\delta$ mainly controls where the pdf of the distribution is placed along the $x$-axis.

\section{Parameter Estimation with Generalized Method of Moments}
As the reader might have noticed, complex isotropic models in general adopt a pdf of integral form, making parameter estimation of which subsequently complicated. Past literature have utilized method of log cumulants (MoLC) \cite{tison2004new} for Heavy-Tailed Rayleigh \cite{achim2006sar} and Markov chain Monte Carlo for Cauchy-Rician \cite{karakucs2022cauchy}, but these prior works have proven futile in front of a more complicated CI$\alpha$SR model. The readers can refer to Appendix \ref{Appendix_failed GeneralizedMoMforCR} for details of these failed efforts. In this section, the thesis would like to propose a quasi-analytical parameter estimator based on generalization of method of moments, despite not completely analytical due to the involvement of a necessary root-finding algorithm, the estimator is otherwise fast and succinct.

The Method of moments requires users to derive the mathematical expectation of a given function, and compare it to the empirical expectation calculated from samples to form:
\begin{equation}
	{\int_{X}^{}{f(x)~p(x)~dx}} = E\left\lbrack {f(x)} \right\rbrack = {\lim\limits_{N\rightarrow\infty}{\sum\limits_{i = 1}^{N}{f\left( x_{i} \right)}}}.
	\label{eq.GeneralizeMoM}
\end{equation}
Conventionally, $f(x)$ are chosen to be power functions due to their mathematical simplicity in the evaluation of the expectation integral on the left-hand side of equation \ref{eq.GeneralizeMoM} and the numerical calculation of the empirical moment from data on the right-hand side of equation \ref{eq.GeneralizeMoM}. However, expectations of power functions for CI$\alpha$SR model have been proven difficult to solve even for the HTR \cite{achim2006sar}. A novel method of Bessel moments (MoBM) is proposed in this work to help resolve the the obstinate parameter estimation problem concerning models derived from the $\alpha$-stable family. We discovered that by assuming $f(x)$ in Eq. \ref{eq.GeneralizeMoM} to be a Bessel function of the first kind, the function conveniently cancels with the pdf term to yield a compact result. The derivation of this Bessel moment for CI$\alpha$SR can be summarized as follows.

By setting the function to $f(x)=J_0 (ax)$, the expectation of said function under a CI$\alpha$SR distribution is
\begin{equation}
	E\left\lbrack {f(x)} \right\rbrack = {\int_{0}^{\infty}{\omega{\exp\left( - \gamma^{\alpha}\omega^{\alpha} \right)}J_{0}\left( {\omega\delta} \right)~\left\lbrack {\int_{0}^{\infty}{x~J_{0}\left( {\omega x} \right)~J_{0}\left( {ax} \right)~dx}} \right\rbrack~d\omega}}.
\end{equation}
By Invoking the following identity \cite{gradshteyn2014table} to integrate out $x$:
\begin{equation}
	{\int_{0}^{\infty}{k~J_{n}\left( {ka} \right)~J_{n}\left( {kb} \right)~dk}} = \frac{1}{a}~{\delta}_{D}\left( {b - a} \right)~~~~~\left\lbrack {n = 0,~1,~2\ldots} \right\rbrack,
\end{equation}
where ${\delta}_{D}\left( {k - k_{0}} \right)$ is the Dirac delta function. This results in
\begin{equation}
	\begin{split}
		E\left\lbrack {J_{0}\left( {ax} \right)} \right\rbrack &= \frac{1}{a}{\int_{0}^{\infty}{\omega{\exp\left( {- \gamma^{\alpha}\omega^{\alpha}} \right)}J_{0}\left( {\omega\delta} \right) {\delta}_{D}\left( {\omega - a} \right)~d\omega}} \\
		&= {\exp\left( {- \gamma^{\alpha}a^{\alpha}} \right)}J_{0}\left( {a\delta} \right).
	\end{split}
\end{equation}
Despite having a Bessel term in its final form, $E\left\lbrack {J_{0}\left( {ax} \right)} \right\rbrack$ as a function of $a$ is obviously well behaved: the first term $\exp\left( {- \gamma^{\alpha}a^{\alpha}} \right)$ is positive and monotonically decreasing; the second term $J_{0}\left( {a\delta} \right)$ as a Bessel function oscillates with an asymptotic period of roughly $\pi/\delta$ as $a$ becomes large, and is also positive and monotonically decreasing before the first root. 
\begin{figure}
	\centering
	\subcaptionbox{\label{Fig_BesselMomentCurve_sub1}}
	{\includegraphics[width=0.32\linewidth]{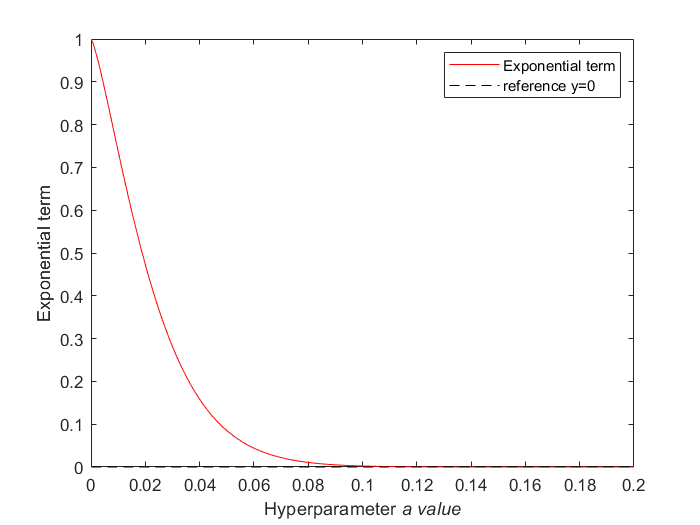}}
	\subcaptionbox{\label{Fig_BesselMomentCurve_sub2}}
	{\includegraphics[width=0.32\linewidth]{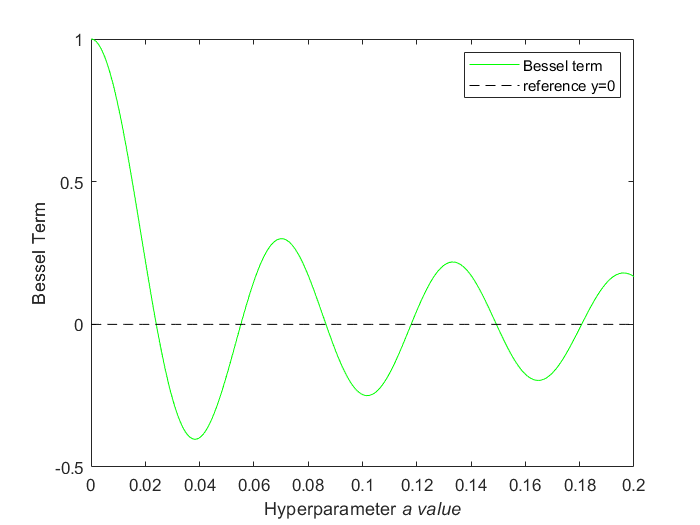}}
	\subcaptionbox{\label{Fig_BesselMomentCurve_sub3}}
	{\includegraphics[width=0.32\linewidth]{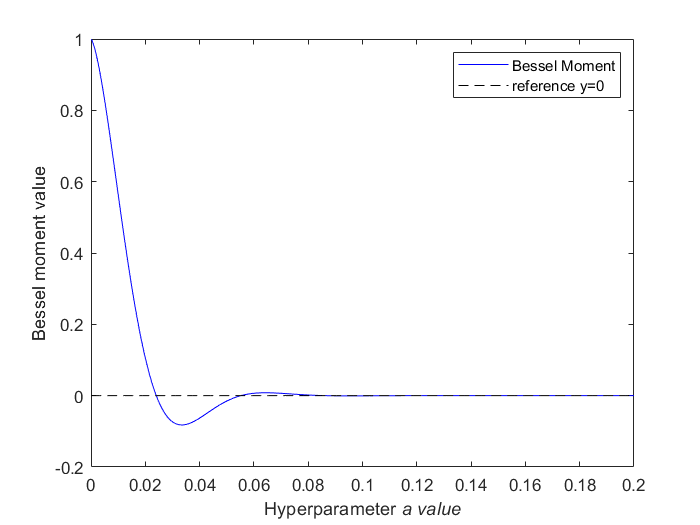}}
	\caption{Shape of the Bessel moment and the two terms contributing to it: (a) plotting of the exponential term; (b) plotting of the Bessel term; (c) plotting of the Bessel moment.}
	\label{Fig_BesselMomentCurve}
\end{figure}

In fact, the shape of $E\left\lbrack {J_{0}\left( {ax} \right)} \right\rbrack$ will behave similar to zeroth-order Bessel function in respect of zeros and signs, but rapidly closing in to the $a$-axis due to the exponential term. Since the roots of Bessel function are readily available with high precision, it is feasible to increase  the value of $a$ in steps until the empirical expectation $\sum{J_{0}\left( {a_{0}x_{i}} \right)}$ goes down to zero, and the delta estimate should equal to the quotient of Bessel first root (approximate value 2.405) by ${a}_{0}$. Another fact to help with finding the root ${a}_{0}$ is that one can hazard a guess of the approximate $\delta$ value, since it is in general of the same magnitude as sample mean, this approximate value can be used as reference for  step of increase in $a$. Pseudo-code for the $\delta$ estimation procedure is presented in 
\begin{algorithm}[H]
	\caption{Pseudo-Code for $\delta$ Estimation}\label{alg:alg1}
	\begin{algorithmic}
		\STATE 
		\STATE {\textsc{initialize}} $a=0$
		\STATE step$=0.01*\frac{1}{N}*{\sum x_{i}}$
		\STATE while ${\sum{J_{0}\left( {ax_{i}} \right)}} > 0$
		\STATE \hspace{0.5cm}$a=a+{\rm step}$
		\STATE end while
		\STATE ${\hat{\delta}}=2.405/a$
	\end{algorithmic}
	\label{alg1}
\end{algorithm}

Once the location parameter estimate $\hat{\delta}$ is acquired, the remaining two parameters can be solved in closed form as follows:
\begin{equation}
	\left\{ \begin{matrix}
		{\hat{\alpha} = {{\ln\left\{ \frac{{\ln{E\left\lbrack {J_{0}\left( {a_{1}x} \right)} \right\rbrack}} - {\ln{J_{0}\left( {a_{1}\delta} \right)}}}{{\ln{E\left\lbrack {J_{0}\left( {a_{2}x} \right)} \right\rbrack}} - {\ln{J_{0}\left( {a_{2}\delta} \right)}}} \right\}}/{\ln\left( \frac{a_{1}}{a_{2}} \right)}}} \\
		{\hat{\gamma} = \frac{1}{a_{3}}\left\{ {\ln\frac{J_{0}\left( {a_{3}\hat{\delta}} \right)}{E\left\lbrack {J_{0}\left( {a_{3}x} \right)} \right\rbrack}} \right\}^{1/\hat{\alpha}~}}, \\
	\end{matrix} \right.
\end{equation}
where $a_{i}~(i=1, 2, 3)$ are arbitrary hyperparameters of different values. During experiments with synthetic CI$\alpha$SR data, it is discovered that they yield best performance when chosen around the magnitude of 0.01. One possible explanation for this choice is that the subsequent value of $a_{i}\delta$ falls around the first root of Bessel function, away from zero so that the slope is large enough and Bessel is sensitive to the change of $\delta$, and not too far away so that Bessel can still be treated as an injective mapping.

Since the estimator offers analytical solutions to parameters $\hat{\alpha}$ and $\hat{\gamma}$, and only involves a simple iterative operation in the acquisition of $\hat{\delta}$, it is computationally economic, and makes a thorough experiment possible on an enormous amount of synthetically generated data as displayed in the following section. The generalization of method of moments to a special function like Bessel function can also inspire the development of alternative generic moment options for solving other analytically complicated distributions.

\section{Experimental Study}
\subsection{Simulations with Synthetic Data}
To prove the robustness of the estimator, the method of Bessel moments estimator is first implemented on synthetic data generated according to the CI$\alpha$SR model. Synthetic data are generated using the following steps: \cite{samorodnitsky2017stable}.
\begin{enumerate}
	\item{	Generate a one-sided $\alpha$-stable random variable  $\left. A \right.\sim S_{\frac{\alpha}{2}}\left( {\left( {\cos\frac{\alpha\pi}{4}} \right)^{\frac{2}{\alpha}},~1,~0} \right)$ using the CMS method \cite{chambers1976method}.
	}
	\item{Generate two i.i.d, zero-mean Gaussian random variables $\left. G_{i} \right.\sim N\left( \mu = 0,~2\gamma^{2} \right)$.
	}
	\item{	Combine the random variables in the following manner $\overset{\rightarrow}{X} = \left( {A^{1/2}G_{1} + \delta_{1},~A^{1/2}G_{2} + \delta_{2}} \right)$.
		The amplitude $X = \left| \overset{\rightarrow}{X} \right|$ is a CI$\alpha$SR random variable.}
\end{enumerate}

\begin{figure}
	\centering
	\subcaptionbox{\label{Fig_DataSizeVSPerformance_sub1}}
	{\includegraphics[width=0.24\linewidth]{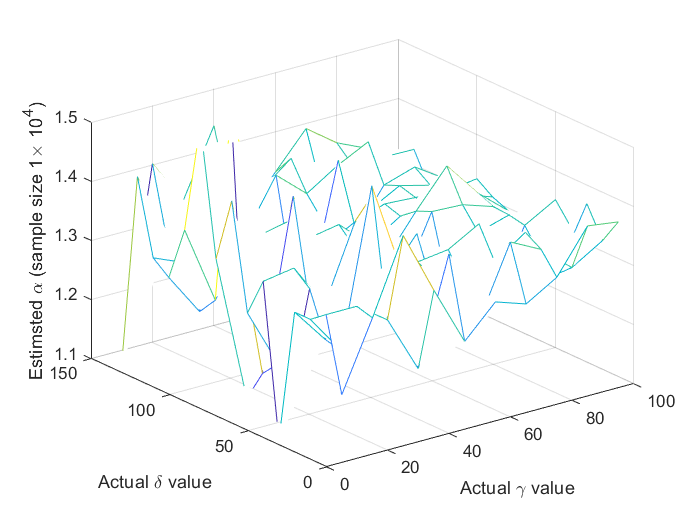}}
	\subcaptionbox{\label{Fig_DataSizeVSPerformance_sub2}}
	{\includegraphics[width=0.24\linewidth]{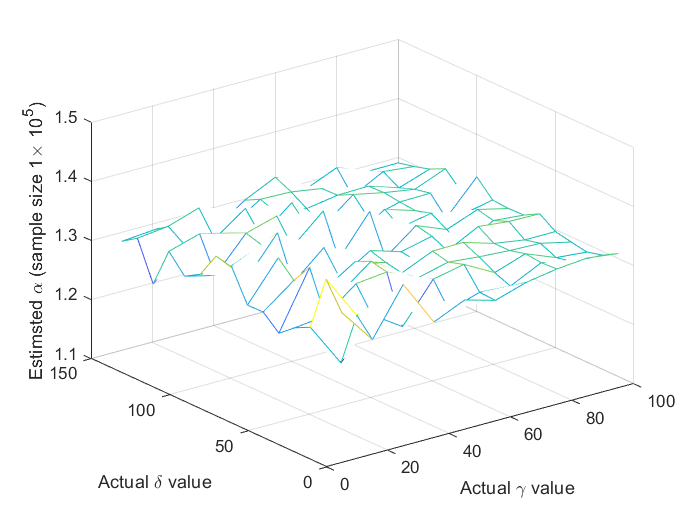}}
	\subcaptionbox{\label{Fig_DataSizeVSPerformance_sub3}}
	{\includegraphics[width=0.24\linewidth]{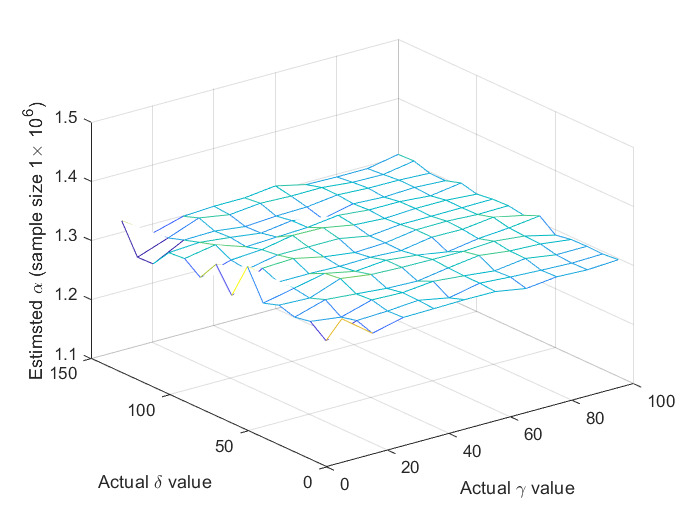}}
	\subcaptionbox{\label{Fig_DataSizeVSPerformance_sub4}}
	{\includegraphics[width=0.24\linewidth]{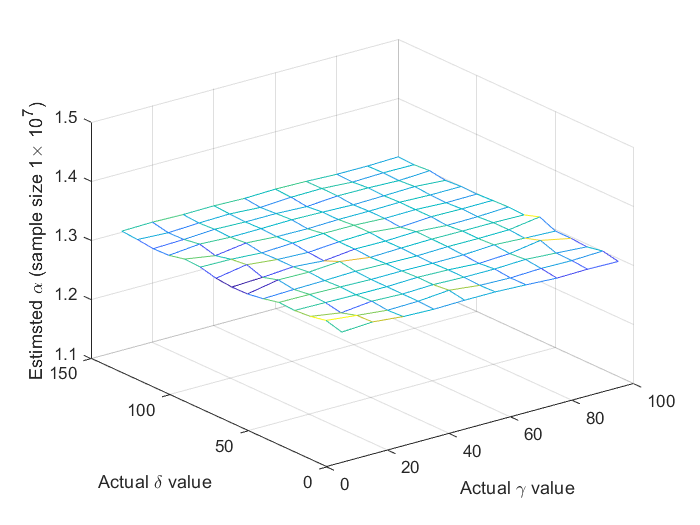}}
	\subcaptionbox{\label{Fig_DataSizeVSPerformance_sub5}}
	{\includegraphics[width=0.24\linewidth]{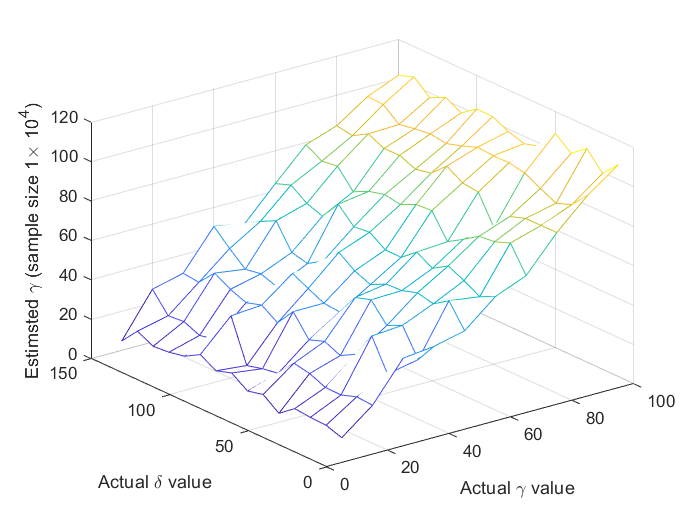}}
	\subcaptionbox{\label{Fig_DataSizeVSPerformance_sub6}}
	{\includegraphics[width=0.24\linewidth]{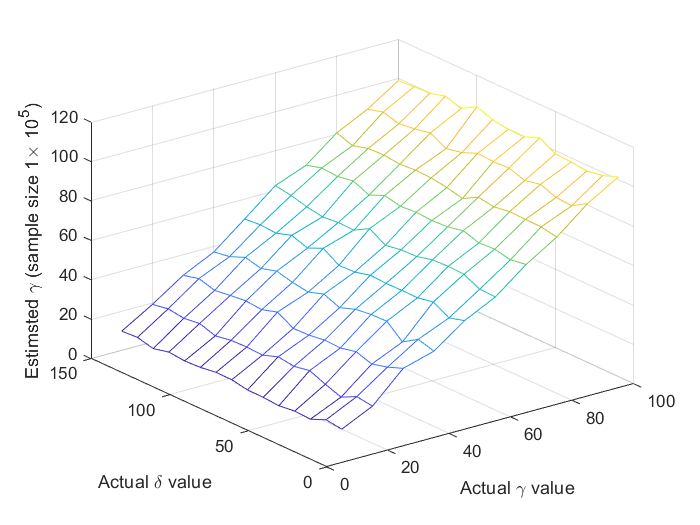}}
	\subcaptionbox{\label{Fig_DataSizeVSPerformance_sub7}}
	{\includegraphics[width=0.24\linewidth]{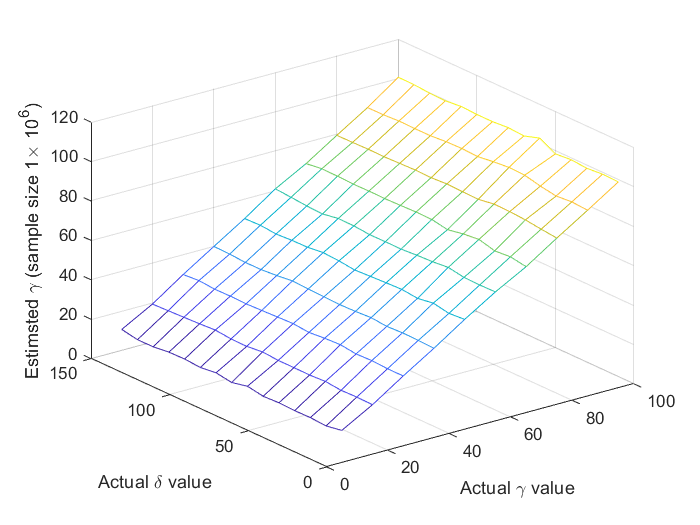}}
	\subcaptionbox{\label{Fig_DataSizeVSPerformance_sub8}}
	{\includegraphics[width=0.24\linewidth]{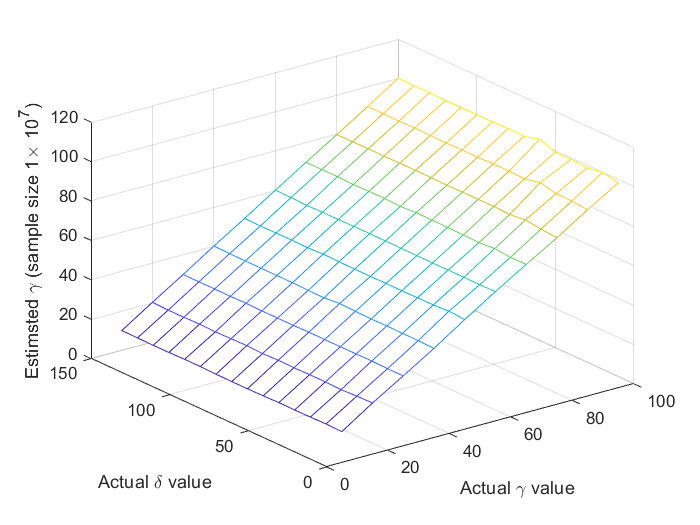}}
	\subcaptionbox{\label{Fig_DataSizeVSPerformance_sub9}}
	{\includegraphics[width=0.24\linewidth]{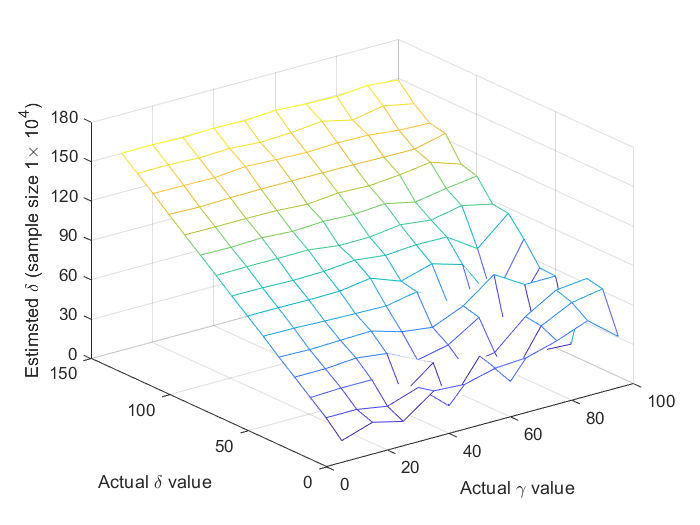}}
	\subcaptionbox{\label{Fig_DataSizeVSPerformance_sub10}}
	{\includegraphics[width=0.24\linewidth]{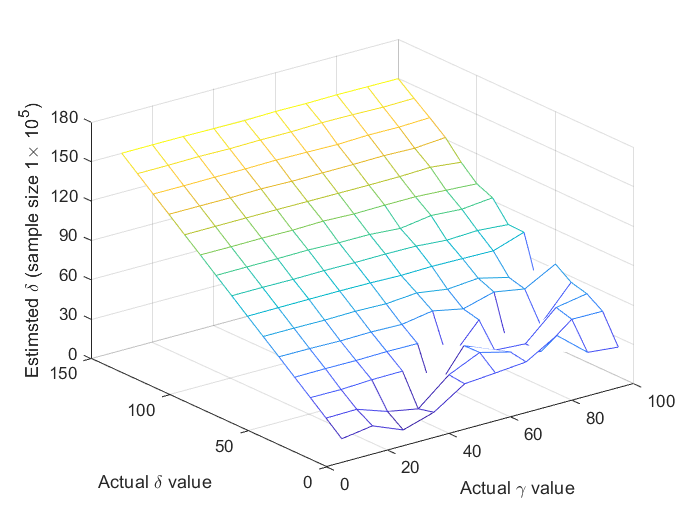}}
	\subcaptionbox{\label{Fig_DataSizeVSPerformance_sub11}}
	{\includegraphics[width=0.24\linewidth]{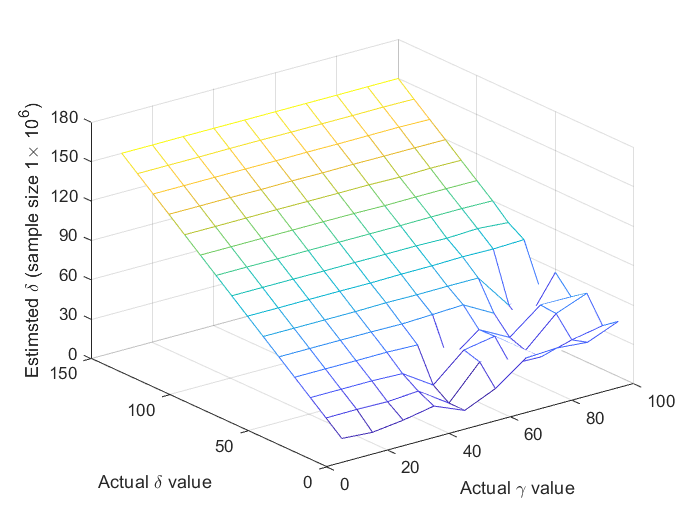}}
	\subcaptionbox{\label{Fig_DataSizeVSPerformance_sub12}}
	{\includegraphics[width=0.24\linewidth]{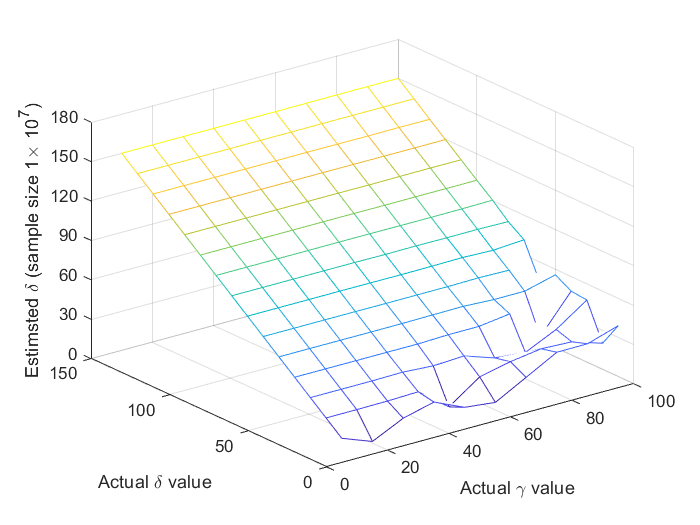}}
	\caption{Test of how size of synthetic sample space affect the stability of MoBM estimator: the X-axis and Y-axis within each subfigure reflects how the estimation is affected by the change of actual $\gamma$ and $\delta$ values during data generation; four columns of subfigures correspond to estimated parameters given synthetic dataset size from $1 \times 10^4$ to $1 \times 10^7$, left to right; the three rows of subfigures corresponds to parameter estimation of $\alpha$, $\gamma$, and $\delta$, respectively.}
	\label{Fig_DataSizeVSPerformance}
\end{figure}
To begin with, a quick experiment is conducted to study the effect of dataset size on the performance of the MoBM estimator. Figure \ref{Fig_DataSizeVSPerformance} incorporates MoBM estimator performance against different synthetic dataset size from $1 \times 10^4$ to $1 \times 10^7$. In general, a smooth surface of the subfigures suggests that the estimation is stable, and the surface being parallel to one or two axes suggests that the estimation of one parameter is independent from the other parameters. Apparently, the stability of estimation of all three parameters are directly dependent on the number of samples, and a larger data set can ensure a more accurate performance of the estimator. The thesis conducts synthetic experiments with a data size of $1 \times 10^6$ per dataset in subsequent sections for a trade-off between computation efficiency and stability.

\begin{figure}
	\centering
	\subcaptionbox{\label{Fig_MoBMonSyntheticData_sub1}}
	{\includegraphics[width=0.32\linewidth]{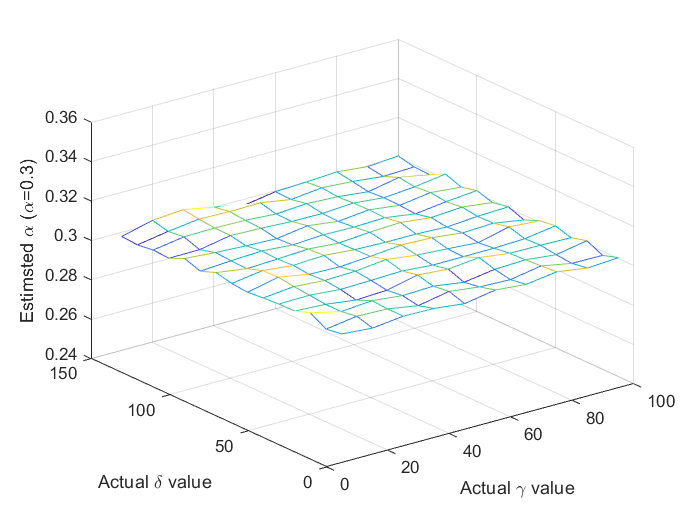}}
	\subcaptionbox{\label{Fig_MoBMonSyntheticData_sub2}}
	{\includegraphics[width=0.32\linewidth]{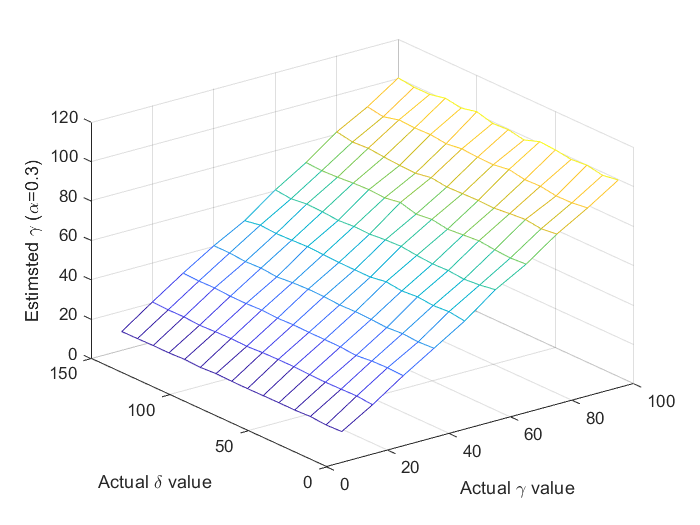}}
	\subcaptionbox{\label{Fig_MoBMonSyntheticData_sub3}}
	{\includegraphics[width=0.32\linewidth]{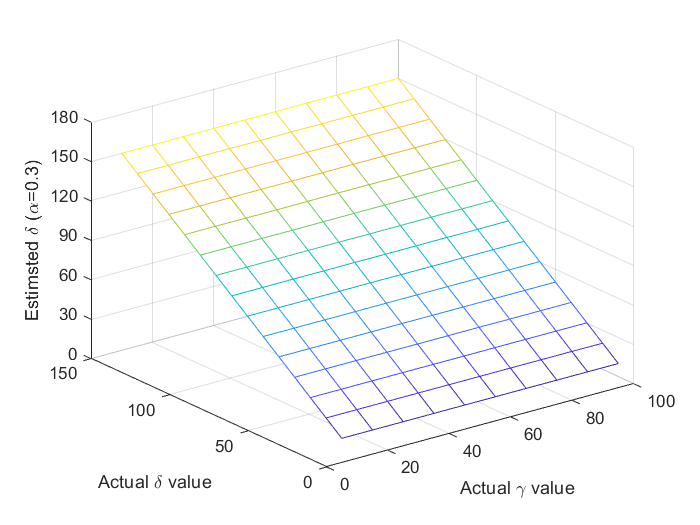}}
	\subcaptionbox{\label{Fig_MoBMonSyntheticData_sub4}}
	{\includegraphics[width=0.32\linewidth]{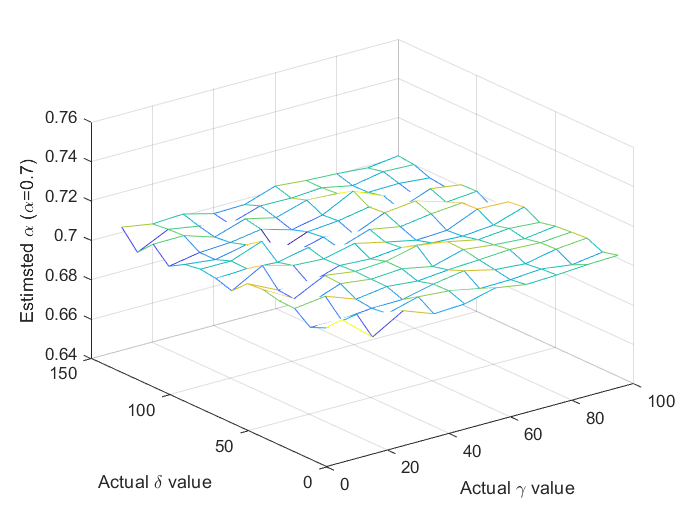}}
	\subcaptionbox{\label{Fig_MoBMonSyntheticData_sub5}}
	{\includegraphics[width=0.32\linewidth]{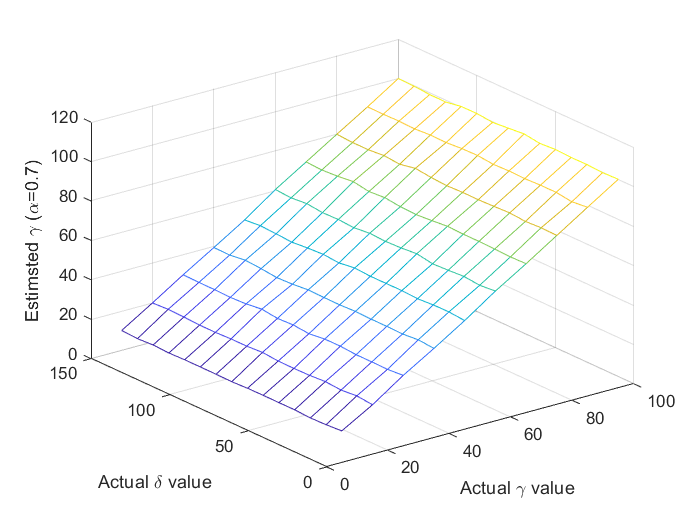}}
	\subcaptionbox{\label{Fig_MoBMonSyntheticData_sub6}}
	{\includegraphics[width=0.32\linewidth]{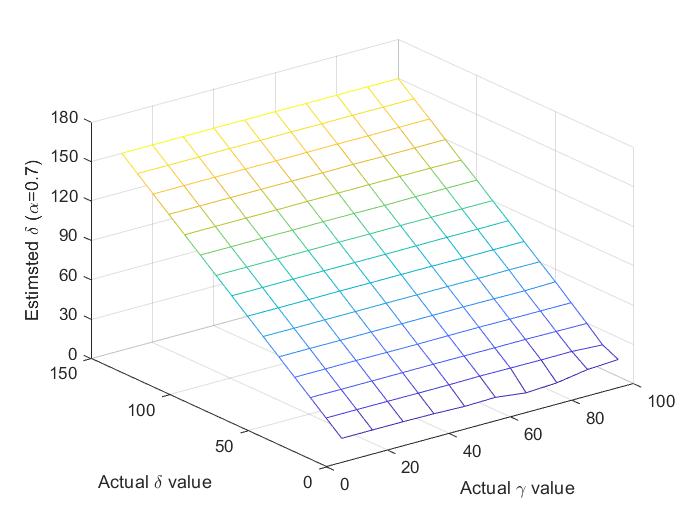}}
	\subcaptionbox{\label{Fig_MoBMonSyntheticData_sub7}}
	{\includegraphics[width=0.32\linewidth]{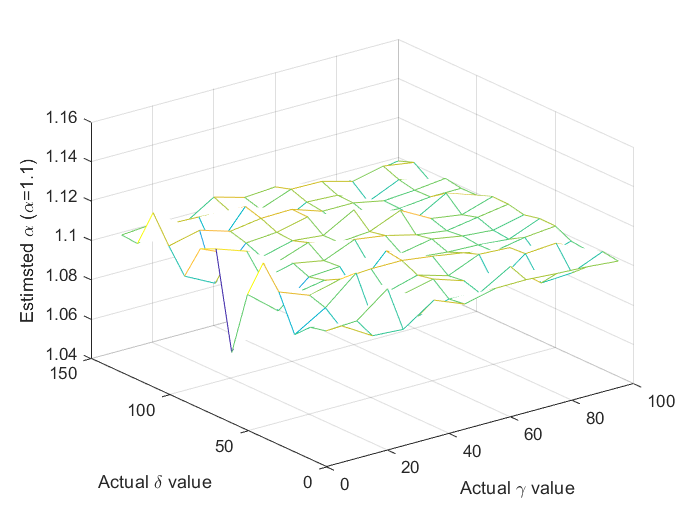}}
	\subcaptionbox{\label{Fig_MoBMonSyntheticData_sub8}}
	{\includegraphics[width=0.32\linewidth]{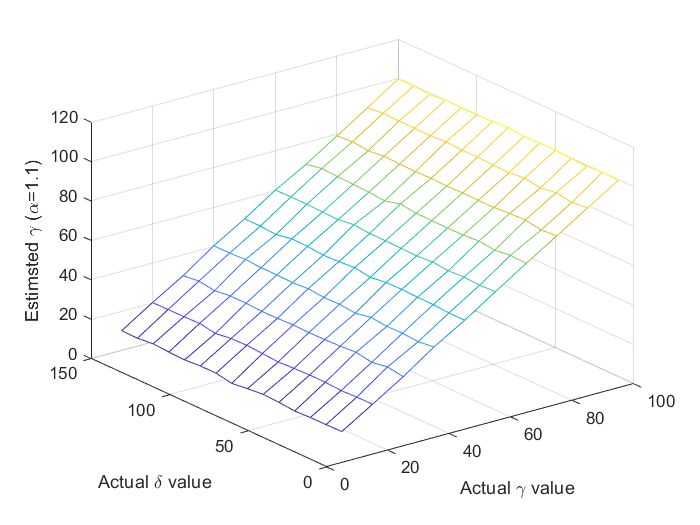}}
	\subcaptionbox{\label{Fig_MoBMonSyntheticData_sub9}}
	{\includegraphics[width=0.32\linewidth]{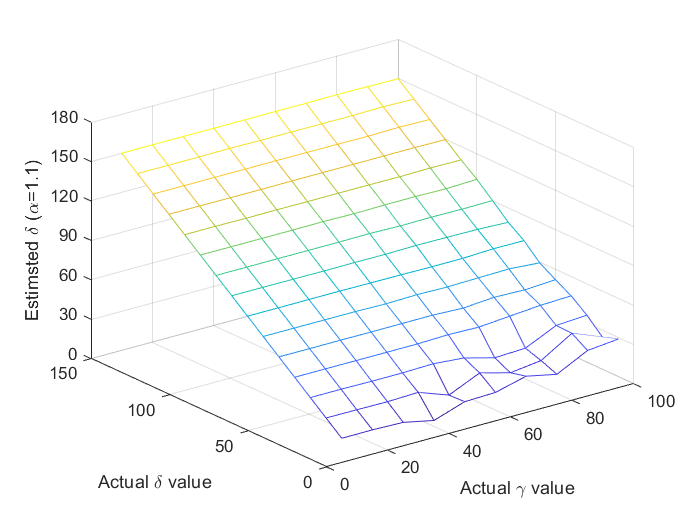}}
	\subcaptionbox{\label{Fig_MoBMonSyntheticData_sub10}}
	{\includegraphics[width=0.32\linewidth]{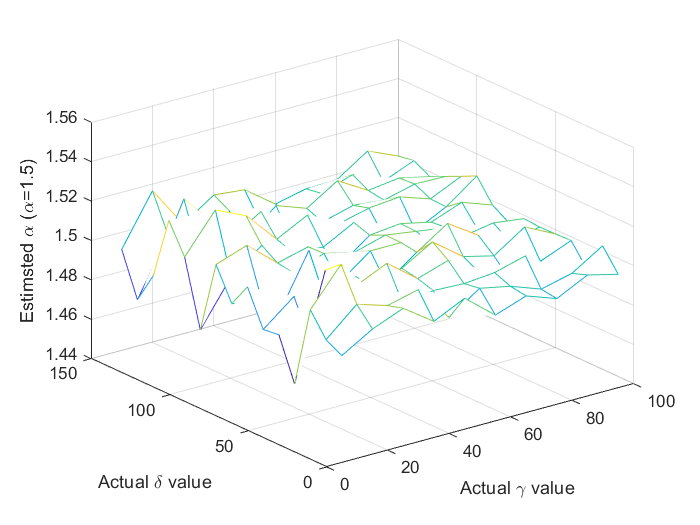}}
	\subcaptionbox{\label{Fig_MoBMonSyntheticData_sub11}}
	{\includegraphics[width=0.32\linewidth]{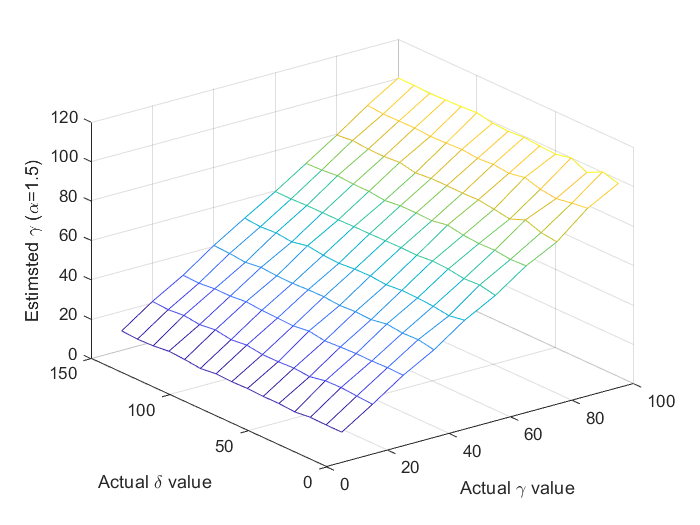}}
	\subcaptionbox{\label{Fig_MoBMonSyntheticData_sub12}}
	{\includegraphics[width=0.32\linewidth]{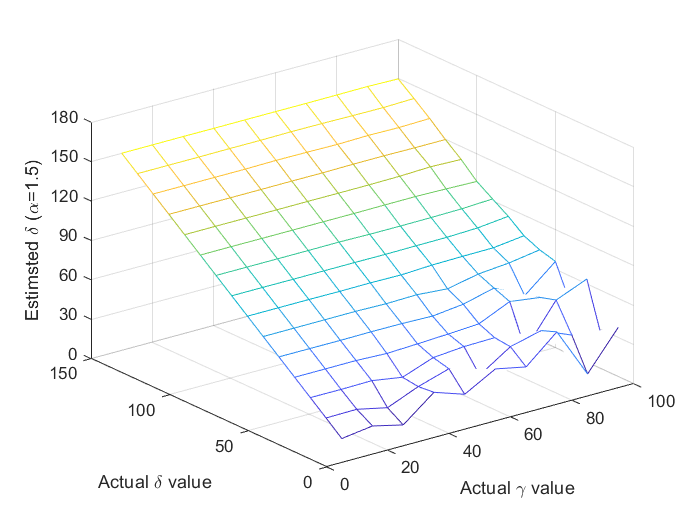}}
	\subcaptionbox{\label{Fig_MoBMonSyntheticData_sub13}}
	{\includegraphics[width=0.32\linewidth]{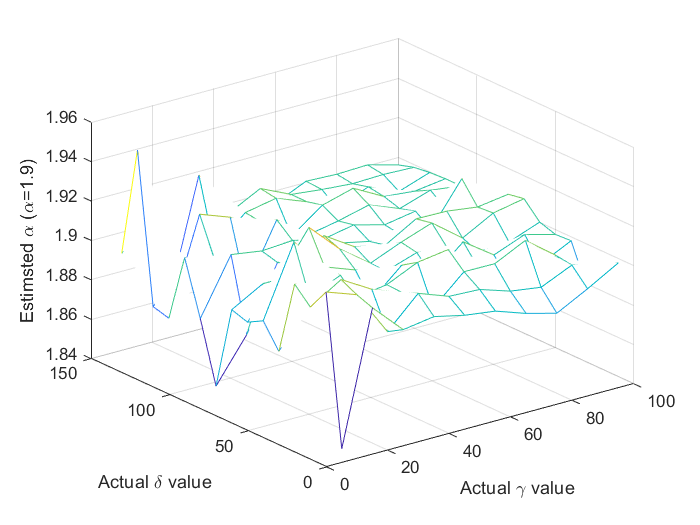}}
	\subcaptionbox{\label{Fig_MoBMonSyntheticData_sub14}}
	{\includegraphics[width=0.32\linewidth]{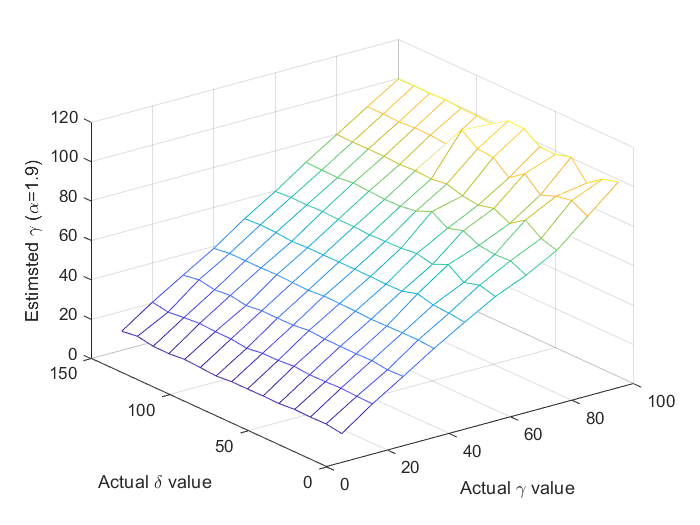}}
	\subcaptionbox{\label{Fig_MoBMonSyntheticData_sub15}}
	{\includegraphics[width=0.32\linewidth]{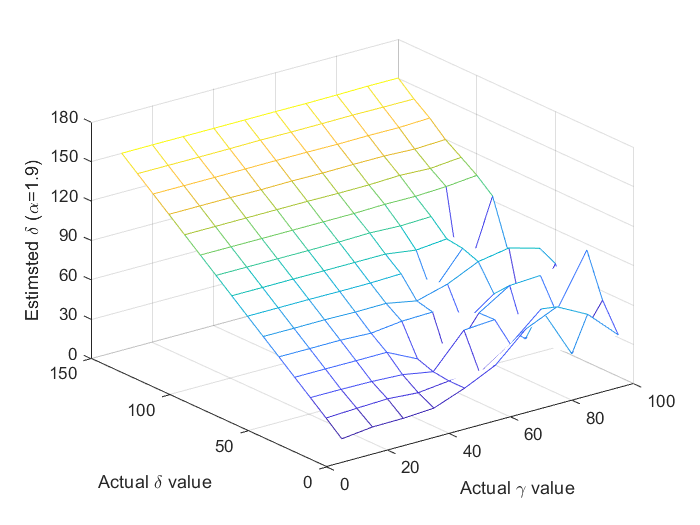}}
	\caption{Estimated parameter value against a comprehensive spectrum of synthetically generated CI$\alpha$SR data with various parameters: the X-axis and Y-axis within each subfigure reflects how the estimation is affected by the change of actual $\gamma$ and $\delta$ values during data generation; three columns of subfigures correspond to estimated $\alpha$, $\gamma$, and $\delta$ parameter from left to right; and five rows of subfigures correspond to synthetic data generated from $\alpha$ value of 0.3, 0.7, 1.1, 1.5, and 1.9 from top to bottom.}
	\label{Fig_MoBMonSyntheticData}
\end{figure}

In this subsection, a total number of 1500 synthetic data sets are generated to cover as wide a range of combination of the three parameters, namely 10 different $\alpha$ from 0.1 to 1.9, by 10 different $\gamma$ from 10 to 100, and by 15 different $\delta$ from 10 to 150. Each synthetic data set contains 1E6 samples to simulate the typical size of a SAR image patch of 1000 by 1000 pixels. For the sake of editorial succinctness, Fig. \ref{Fig_MoBMonSyntheticData} displays only the results when $\alpha$ is at 0.3, 0.7, 1.1, 1.5, and 1.9.

\begin{figure}
	\centering
	\subcaptionbox{\label{Fig_MSEandDklforSyntheticData_sub1}}
	{\includegraphics[width=0.45\linewidth]{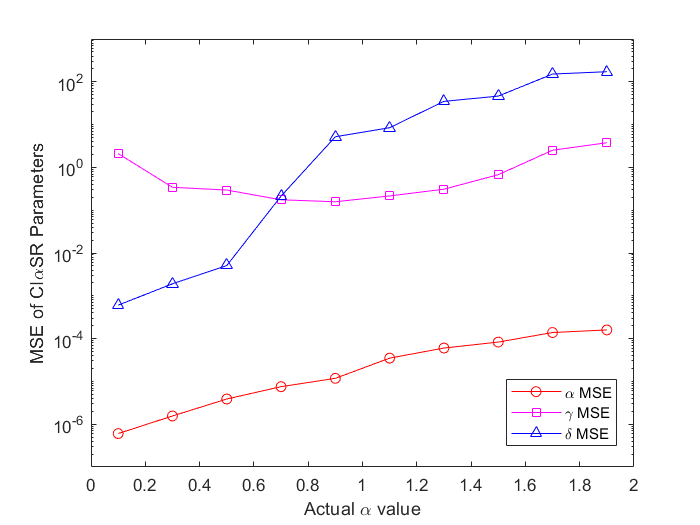}}
	\subcaptionbox{\label{Fig_MSEandDklforSyntheticData_sub2}}
	{\includegraphics[width=0.45\linewidth]{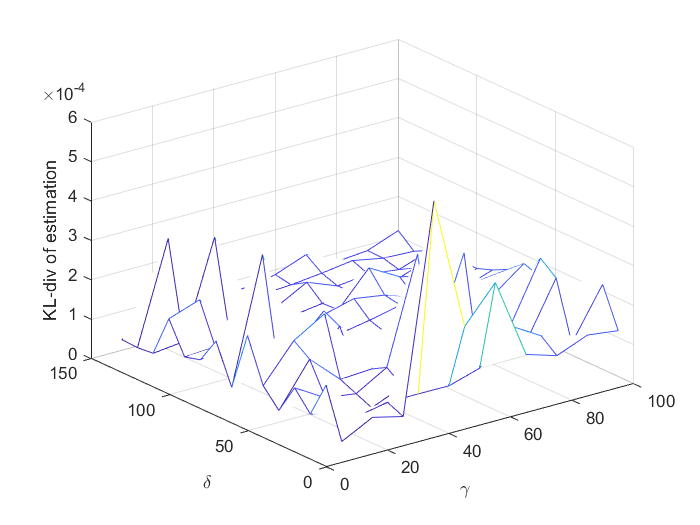}}
	\caption{Performance metrics for the synthetic data experiment: (a) MSE of estimated parameters according to varying $\alpha$ parameter value. (b) KL=div of estimation when $\alpha$=1.9.}
	\label{Fig_MSEandDklforSyntheticData}
\end{figure}

A quick comparison between the sub-figures in Fig. \ref{Fig_MoBMonSyntheticData} can offer readers with a straight-forward reference on performance of the Bessel moment estimator. The estimation of $\alpha$ fluctuates as actual $\alpha$ value used in data generation gets higher, and is most sensitive to an extremely small $\gamma$ value. The estimation of $\gamma$ also experiences minor influence as actual $\alpha$ gets higher or extremely low ($\alpha$ close to $0$, which does not occur in any practical scenario), but the effect is negligible; the estimation of $\delta$ parameter is the most problematic, since it becomes unstable when actual $\alpha$ and $\gamma$ is large while actual $\delta$ is small. This is most probably because high $\alpha$ and $\gamma$ leads to a homogeneous, dispersed data histogram, and the proposed estimator lacks adequate ability to distinguish this dispersed feature from a data set of large shift. Fig. \ref{Fig_MSEandDklforSyntheticData} includes some of the performance metrics gathered in the synthetic data experiment such as mean square error (MSE) and Kullback-Leibler divergence (KL-div) to evaluate the estimator quantitatively. Fig. \ref{Fig_MSEandDklforSyntheticData_sub1} shows the MSE of the three estimated parameters as $\alpha$ range from its domain of $(0, 2]$. Fig. \ref{Fig_MSEandDklforSyntheticData} (b) shows particularly the KL-div of the $\alpha$=1.9 case, when the estimation accuracy is affected the most.

The MSE performance generally shows the tendency of estimation becoming unstable at large $\alpha$, the only exception being when the MSE of $\gamma$ parameter increases as $\alpha$ approaches zero. A main reason for this phenomenon is that the estimator tends to confuse the dispersion of the pdf with the strong heterogeneity characterized with low $\alpha$ values.
The KL-div results also confirm the fact that estimation becomes unstable when $\gamma$ or $\delta$ is very small. One intriguing fact is that despite $\alpha$ and $\gamma$ parameters are estimated on the basis of $\delta$, the deterioration of $\delta$ estimation at high $\alpha$ and $\gamma$ does not seem to drastically contaminate the two parameters or significantly encumber the overall fitting performance, since the KL-div remains in the magnitude of $1\times{10}^{-4}$. One possible explanation for this is that a less accurate $\delta$ estimation is compensated by an overestimated dispersion, subsequently reducing its damage to the overall fitting performance. Still, Fig. \ref{Fig_MSEandDklforSyntheticData_sub2} and Fig. \ref{Fig_MoBMonSyntheticData_sub15} suggest that goodness-of fit is essentially affected by the inaccurate estimation of the characteristic exponent $\alpha$ and location parameter $\delta$.

\subsection{Experiments on Real Data}
In this section, the proposed model and estimator are implemented on SAR image data of various heterogeneity. Images used in this section are captured by ESA’s C-band SAR satellite Sentinel-1, and include scenes such as desert, vegetation, mountain, sea with ships (w/Sea), and urban areas. Performance of the model is compared with the state-of-the-art models including Weibull, log-normal, $\mathcal{G}^{0}$ \cite{frery1997model} and GGR \cite{karakucs2021generalized}, and is evaluated using Kullback-Leibler divergence (KL-div) and Kolmogorov-Smirnov score (KS-score). Experimental results are shown in the following table with winner models in bold font. There are certain cases where the winning model by KL-div or KS-score standards are not the same, this is because of a draw between the CI$\alpha$SR and GGR model since both are good at modelling moderately heterogeneous data.
\begin{table*}[htbp]
	\centering
	\caption{Performance of Different Models on Various SAR Data}
	\resizebox{1\columnwidth}{!}{
	\begin{tabular}{c c|c c c c c | c c c c c}
		\hline
		\multirow{2}{*}{No.} & \multirow{2}{*}{Scene} & \multicolumn{5}{c|}{KL-div} & \multicolumn{5}{c}{KS-score} \\
		& & CI$\alpha$SR & Weibull & log-normal & $\mathcal{G}^{0}$ & GGR & CI$\alpha$SR & Weibull & log-normal & $\mathcal{G}^{0}$ & GGR\\
		\hline
		01 & Desert1 & 0.1286 & 0.1276 & 0.2334 & 0.3337 & \textbf{0.0237} & 0.1077 & 0.1066 & 0.1575 & 0.2938 & \textbf{0.0311} \\
		02 & Desert2 & 0.0750 & 0.2545 & 0.3050 & 0.1446 & \textbf{0.0322} & 0.0988 & 0.1557 & 0.1860 & 0.1659 & \textbf{0.0412} \\
		03 & Desert3 & 0.1079 & 0.1682 & 0.2250 & 0.3910 & \textbf{0.0333} & 0.0926 & 0.1156 & 0.1535 & 0.3124 & \textbf{0.0347} \\
		04 & Vegetation1 & 0.0868 & 0.0967 & 0.2285 & 0.5365 & \textbf{0.0662} & \textbf{0.0688} & 0.0910 & 0.1573 & 0.3603 & 0.1010 \\
		05 & Vegetation2 & 0.1042 & 0.1604 & 0.2139 & 0.3958 & \textbf{0.0220} & 0.0805 & 0.1008 & 0.1527 & 0.3109 & \textbf{0.0268} \\
		06 & Vegetation3 & 0.1246 & 0.1785 & 0.2281 & 0.3577 & \textbf{0.0198} & 0.0999 & 0.1046 & 0.1574 & 0.3012 & \textbf{0.0270} \\
		07 & Mountain1 & 0.1311 & 0.2273 & 0.2871 & 0.1535 & \textbf{0.0523} & 0.1636 & 0.1333 & 0.1800 & 0.1844 & \textbf{0.0898} \\
		08 & Mountain2 & 0.1661 & 0.2335 & 0.2972 & 0.1507 & \textbf{0.0341} & 0.1855 & 0.1358 & 0.1825 & 0.1779 & \textbf{0.0641} \\
		09 & Mountain3 & 0.1601 & 0.2092 & 0.2684 & 0.1889 & \textbf{0.0202} & 0.1631 & 0.1266 & 0.1761 & 0.2102 & \textbf{0.0224} \\
		10 & w/Sea1 & \textbf{0.0164} & 0.1059 & 0.1643 & 0.7558 & 0.1838 & \textbf{0.0371} & 0.1380 & 0.1247 & 0.4445 & 0.2309 \\
		11 & w/Sea2 & \textbf{0.1470} & 0.1882 & 0.2256 & 0.6821 & 0.1672 & \textbf{0.1127} & 0.1750 & 0.1163 & 0.4203 & 0.1962 \\
		12 & w/Sea3 & \textbf{0.0326} & 0.0567 & 0.1694 & 0.5907 & 0.5194 & \textbf{0.0379} & 0.0722 & 0.1362 & 0.3595 & 0.4059 \\
		13 & Urban1 & \textbf{0.0348} & 0.1156 & 0.2072 & 0.3938 & 0.8145 & \textbf{0.0568} & 0.1071 & 0.1362 & 0.3157 & 0.5599 \\
		14 & Urban2 & \textbf{0.0672} & 0.2034 & 0.2230 & 0.5276 & 0.0790 & \textbf{0.0758} & 0.1491 & 0.1278 & 0.3596 & 0.1094 \\
		15 & Urban3 & 0.0725 & 0.1838 & 0.2314 & 0.4939 & \textbf{0.0563} & \textbf{0.0870} & 0.1412 & 0.1287 & 0.3515 & 0.0902 \\
		16 & Urban4 & \textbf{0.0945} & 0.2107 & 0.2712 & 0.3019 & 0.2332 & \textbf{0.1049} & 0.1407 & 0.1542 & 0.2788 & 0.1997 \\
		17 & Urban5 & \textbf{0.0971} & 0.1925 & 0.2568 & 0.2925 & 0.2386 & \textbf{0.0873} & 0.1401 & 0.1641 & 0.2645 & 0.1937 \\
		18 & Urban6 & \textbf{0.0292} & 0.1543 & 0.2416 & 0.2857 & 0.2665 & \textbf{0.0517} & 0.1283 & 0.1507 & 0.2713 & 0.2023 \\
		19 & Urban7 & \textbf{0.0478} & 0.1037 & 0.2157 & 0.2534 & 0.0790 & 0.0728 & 0.1089 & 0.1548 & 0.2488 & \textbf{0.0629} \\
		20 & Urban8 & \textbf{0.0409} & 0.1164 & 0.2267 & 0.1966 & 0.0841 & \textbf{0.0705} & 0.1181 & 0.1602 & 0.2223 & 0.0732 \\
		21 & Urban9 & 0.0596 & 0.1809 & 0.2205 & 0.4205 & \textbf{0.0564} & \textbf{0.0690} & 0.1268 & 0.1381 & 0.3156 & 0.0919 \\
		\hline
	\end{tabular}
        }  
	\begin{tablenotes}
		\item w/Sea: SAR image of sea surface with ships.
	\end{tablenotes}
	\label{Table_CIαSRperformance}
\end{table*}

Performance metrics in Table \ref{Table_CIαSRperformance} show that the proposed CI$\alpha$SR model excels in modelling urban and portal sea data with almost unanimous advantage, this observation coincides with CI$\alpha$SR's theoretical foundation of GCLT which states that the data population of these scenes are to have a heavy power tail. We attribute the superiority of the model at modelling heterogeneous data to the comprehensiveness of it having three independent parameters. The estimated $\alpha$ in heterogeneous data are generally between 1.3 to 1.9, which make amends for the Cauchy-Rician model ($\alpha$=1) for assuming the heterogeneity way higher that the actual SAR image; the estimated $\delta$ is distributed between 20 to 140, which in turn resolves the problem of dominant reflectors neglected by the Heavy-Tailed Rayleigh model. However, CI$\alpha$SR often fails or comes in second position when modelling homogeneous data such as images of vegetation or desert, despite its theoretical capability of modelling data of a wide range of heterogeneity by changing the characteristic exponent $\alpha$.
\begin{figure}
	\centering
	\subcaptionbox{\label{Fig_goodness-of-fitCIαSR_sub1}}
	{\includegraphics[width=0.39\linewidth]{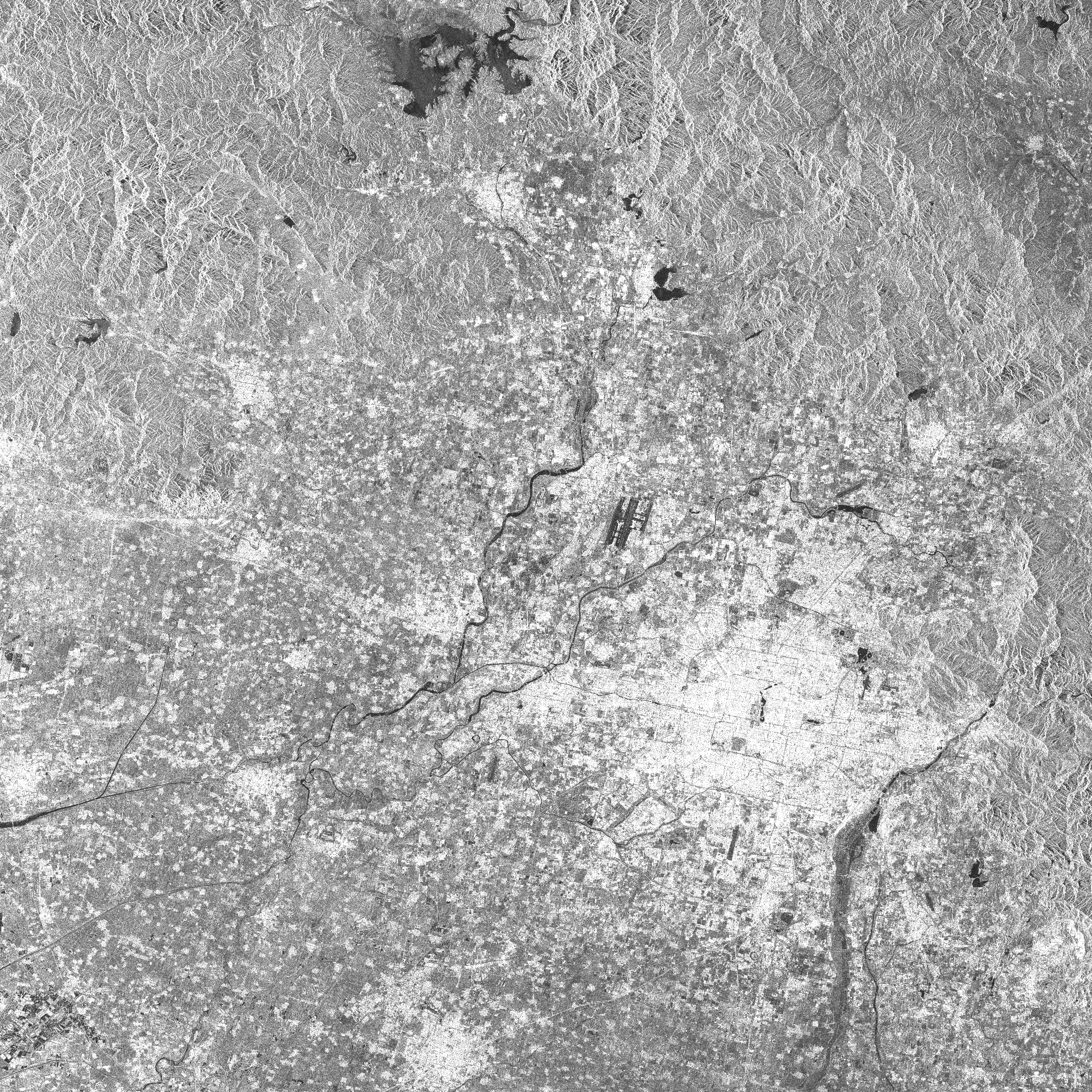}}
	\subcaptionbox{\label{Fig_goodness-of-fitCIαSR_sub2}}
	{\includegraphics[width=0.56\linewidth]{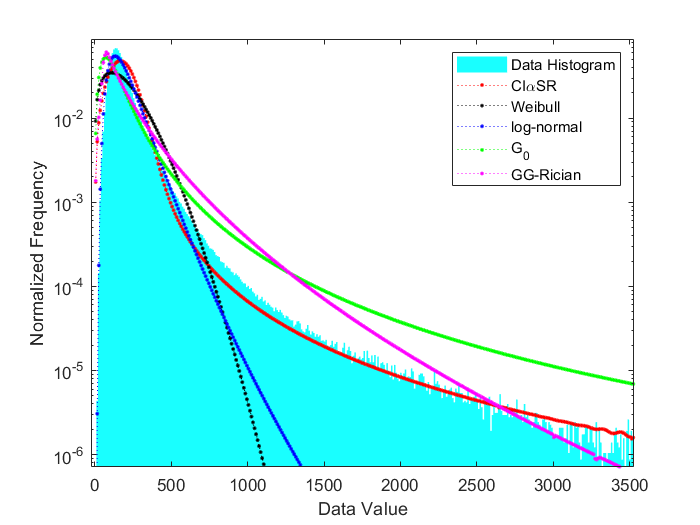}}
	\subcaptionbox{\label{Fig_goodness-of-fitCIαSR_sub3}}
	{\includegraphics[width=0.39\linewidth]{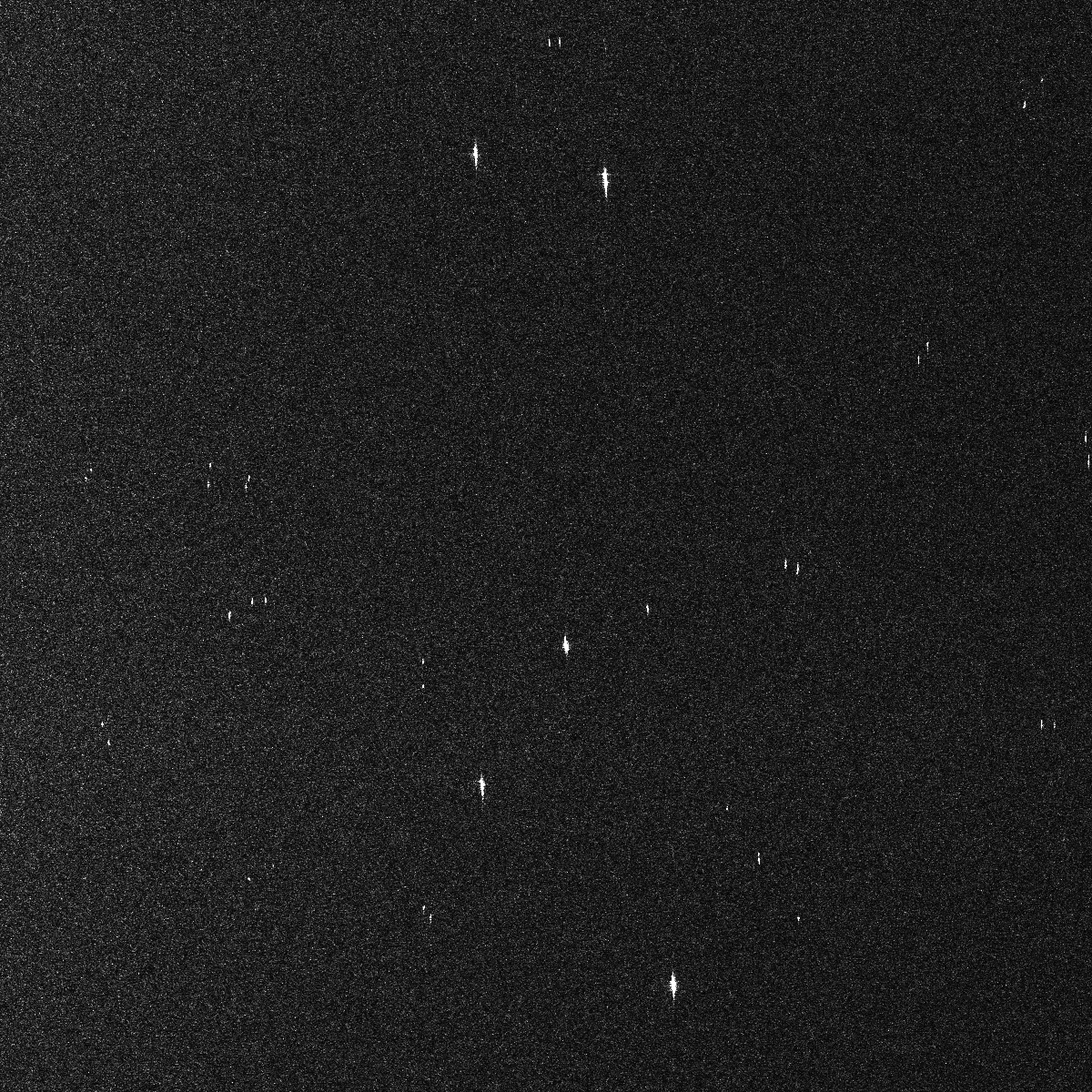}}
	\subcaptionbox{\label{Fig_goodness-of-fitCIαSR_sub4}}
	{\includegraphics[width=0.56\linewidth]{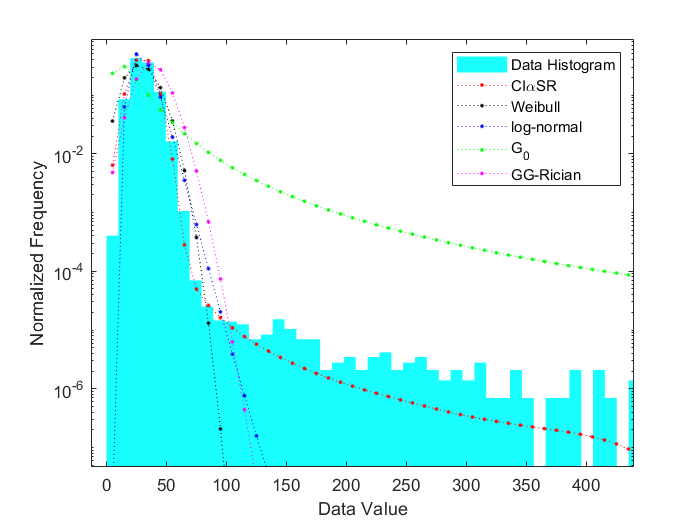}}
	\subcaptionbox{\label{Fig_goodness-of-fitCIαSR_sub5}}
	{\includegraphics[width=0.39\linewidth]{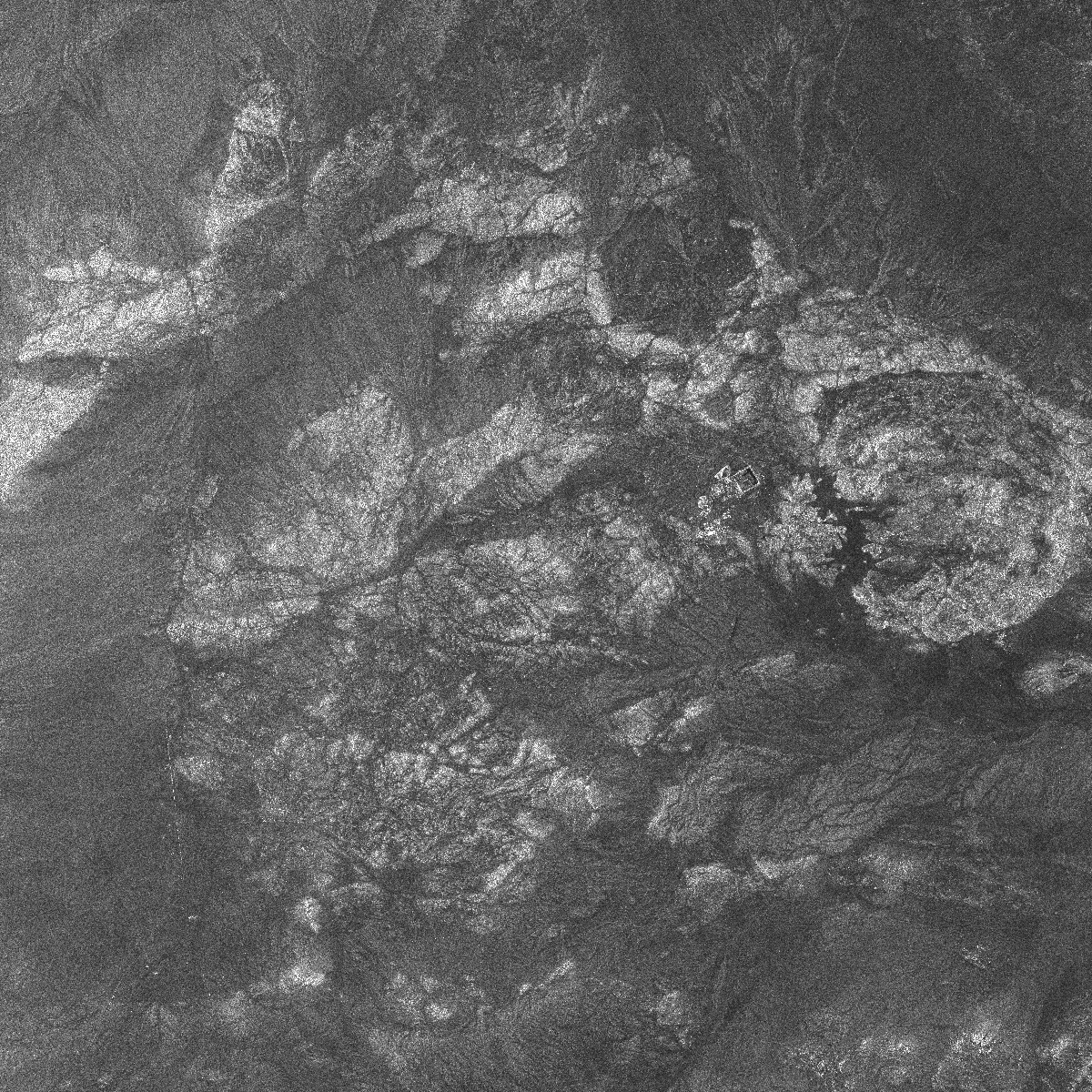}}
	\subcaptionbox{\label{Fig_goodness-of-fitCIαSR_sub6}}
	{\includegraphics[width=0.56\linewidth]{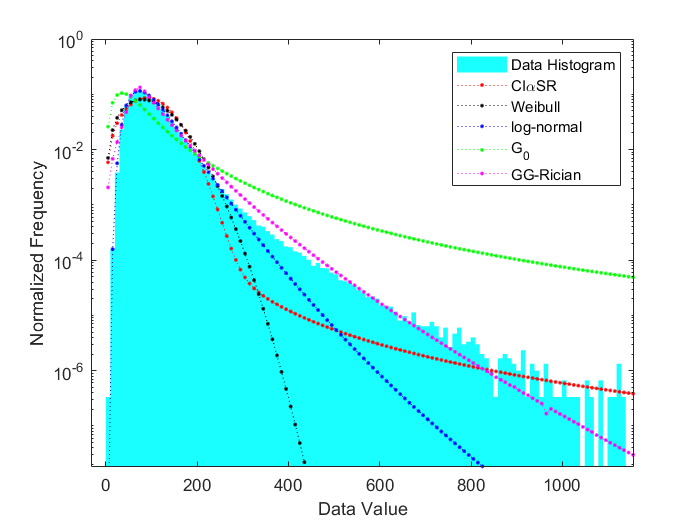}}
	\caption{SAR images and histogram versus candidate model fits of (a, b) Urban 4; (c, d) w/Sea 1; and (e, f) Vegetation 2}
	\label{Fig_goodness-of-fitCIαSR}
\end{figure}

To further study the reasons behind fitting performance of CI$\alpha$SR model, the pdf curves of these candidate models are plotted alongside the SAR image data histogram to provide a direct demonstration of goodness-of-fit. Three sets of image data, namely Urban4, w/Sea2, and Vegetation3 from Table \ref{Table_CIαSRperformance} are selected to accommodate different heterogeneity. According to Figure \ref{Fig_goodness-of-fitCIαSR}, CI$\alpha$SR exemplifies best representation of heavy histogram tails in all three different scenes, but results in Table \ref{Table_CIαSRperformance} shows that the model only wins in Urban4 and w/Sea2 and not in Vegetation3. A closer look into the SAR image histograms may give the answer: while CI$\alpha$SR curve serves as accurate fitting throughout the data histogram in Figure \ref{Fig_goodness-of-fitCIαSR_sub2} and \ref{Fig_goodness-of-fitCIαSR_sub4}, the model fails to accommodate a large part of the transit area between peak and tail (x from 220 to 800) in sub-figure (f), the distinction between bright and dark pixels in sub-figure (a-c) also suggests that Vegetation3 is of less heterogeneous texture than the other two images. One explanation is that the CI$\alpha$SR model tends to sacrifice accurate fitting in the peak-tail transition area to exchange for better representation of the histogram tail, yet this preference drags down its rating in performance metrics such as KL-div and KS-score. One possible solution is to further generalize the model by adding skewness parameter $\beta$ to the model for further flexibility of model pdf shape \cite{kuruoglu2003skewed}, which requires an adequate parameter estimator that the authors wish to address in their future work.

\subsection{Interpretation of Model Parameters}
\begin{figure}
	\centering
	\subcaptionbox{\label{Fig_segmentCIαSR_sub1}}
	{\includegraphics[width=0.49\linewidth]{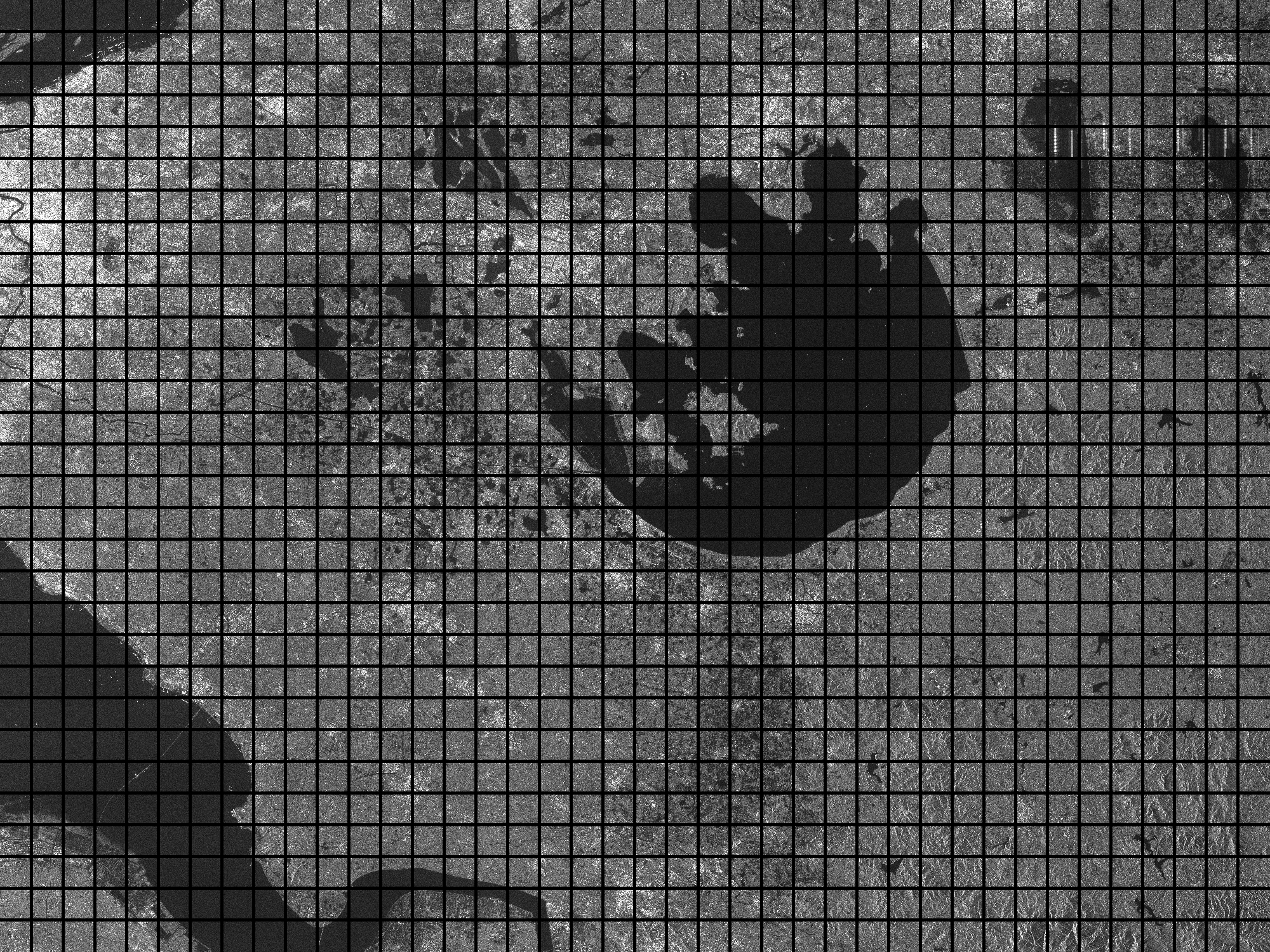}}
	\subcaptionbox{\label{Fig_segmentCIαSR_sub2}}
	{\includegraphics[width=0.49\linewidth]{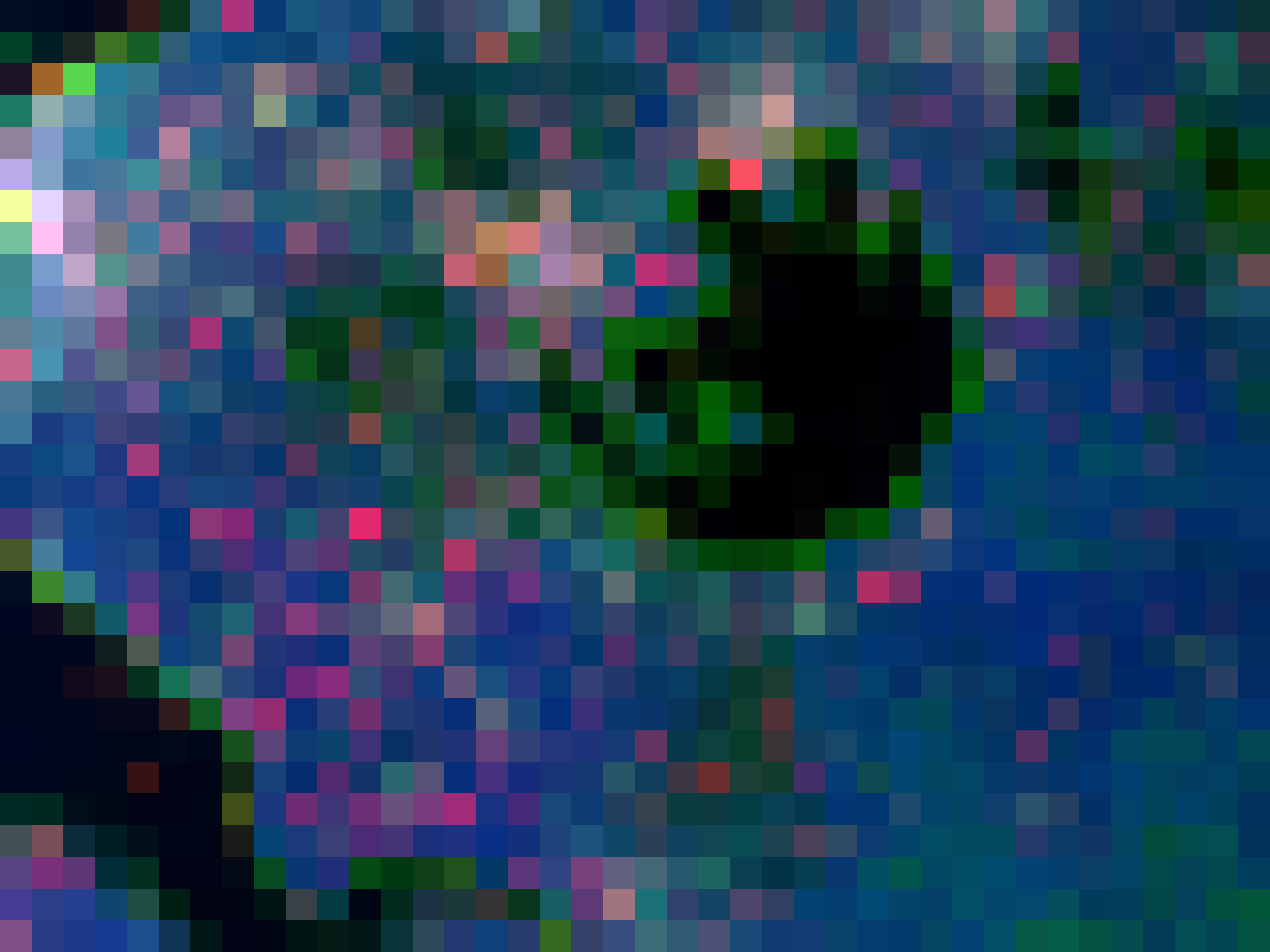}}
	\subcaptionbox{\label{Fig_segmentCIαSR_sub3}}
	{\includegraphics[width=0.32\linewidth]{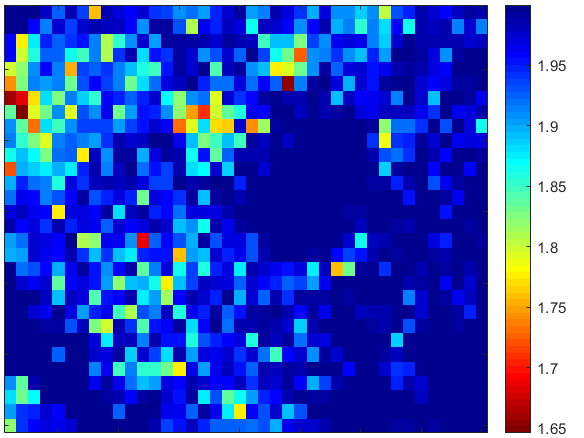}}
	\subcaptionbox{\label{Fig_segmentCIαSR_sub4}}
	{\includegraphics[width=0.32\linewidth]{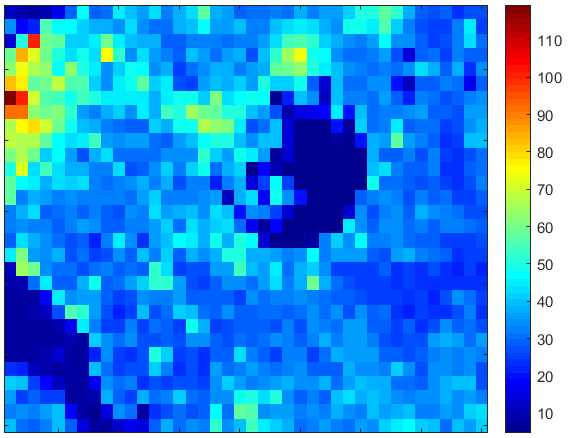}}
	\subcaptionbox{\label{Fig_segmentCIαSR_sub5}}
	{\includegraphics[width=0.32\linewidth]{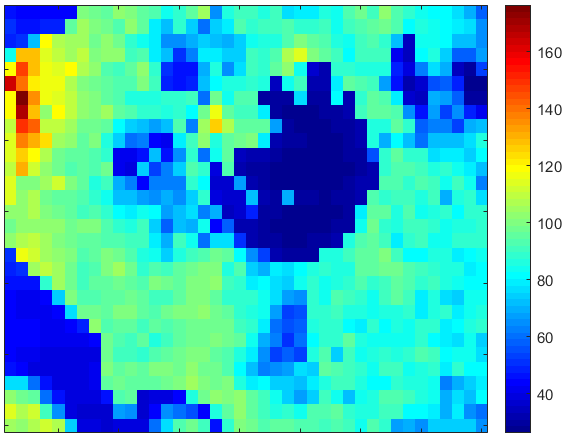}}
	\subcaptionbox{\label{Fig_segmentCIαSR_sub6}}
	{\includegraphics[width=0.49\linewidth]{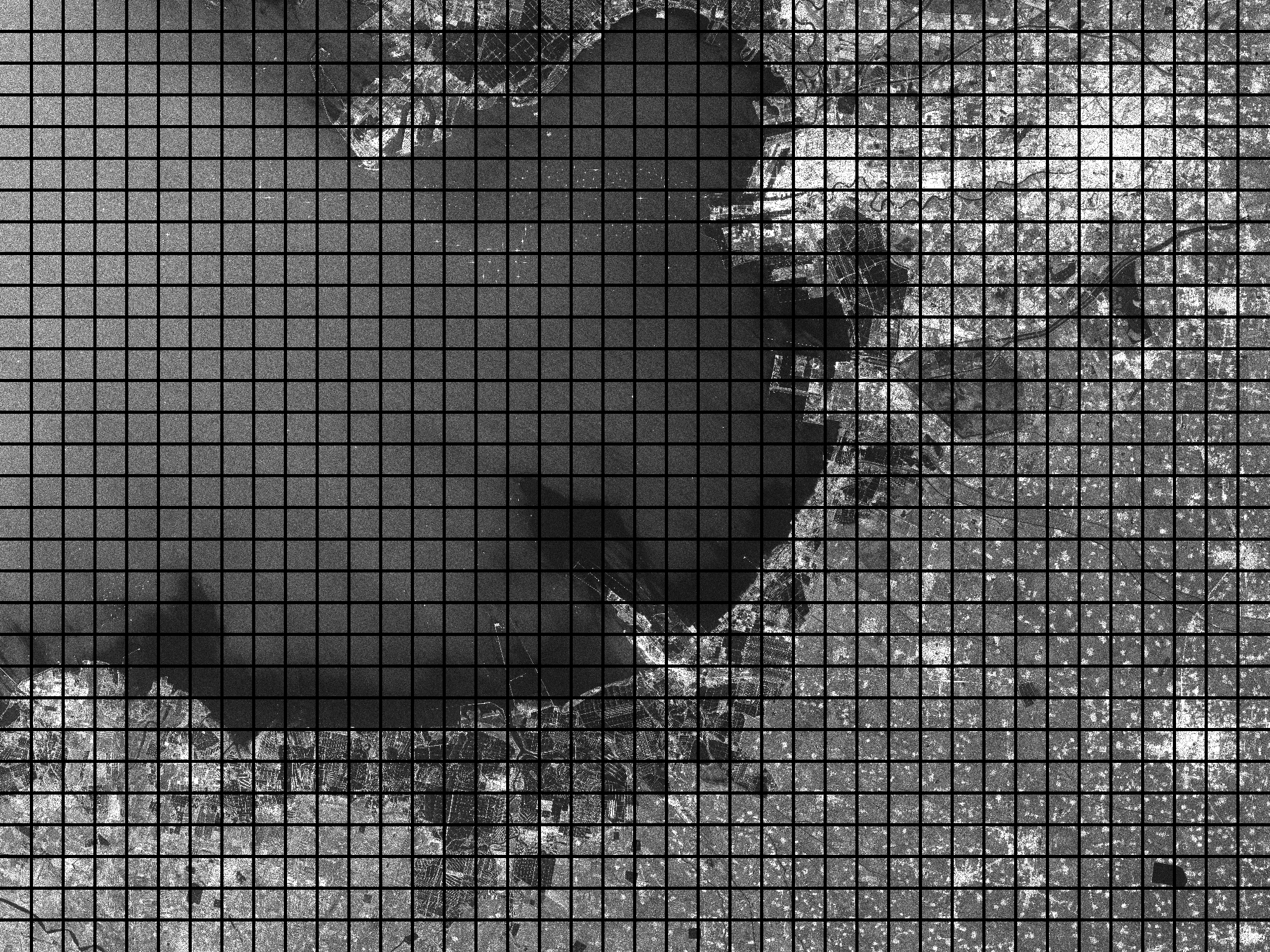}}
	\subcaptionbox{\label{Fig_segmentCIαSR_sub7}}
	{\includegraphics[width=0.49\linewidth]{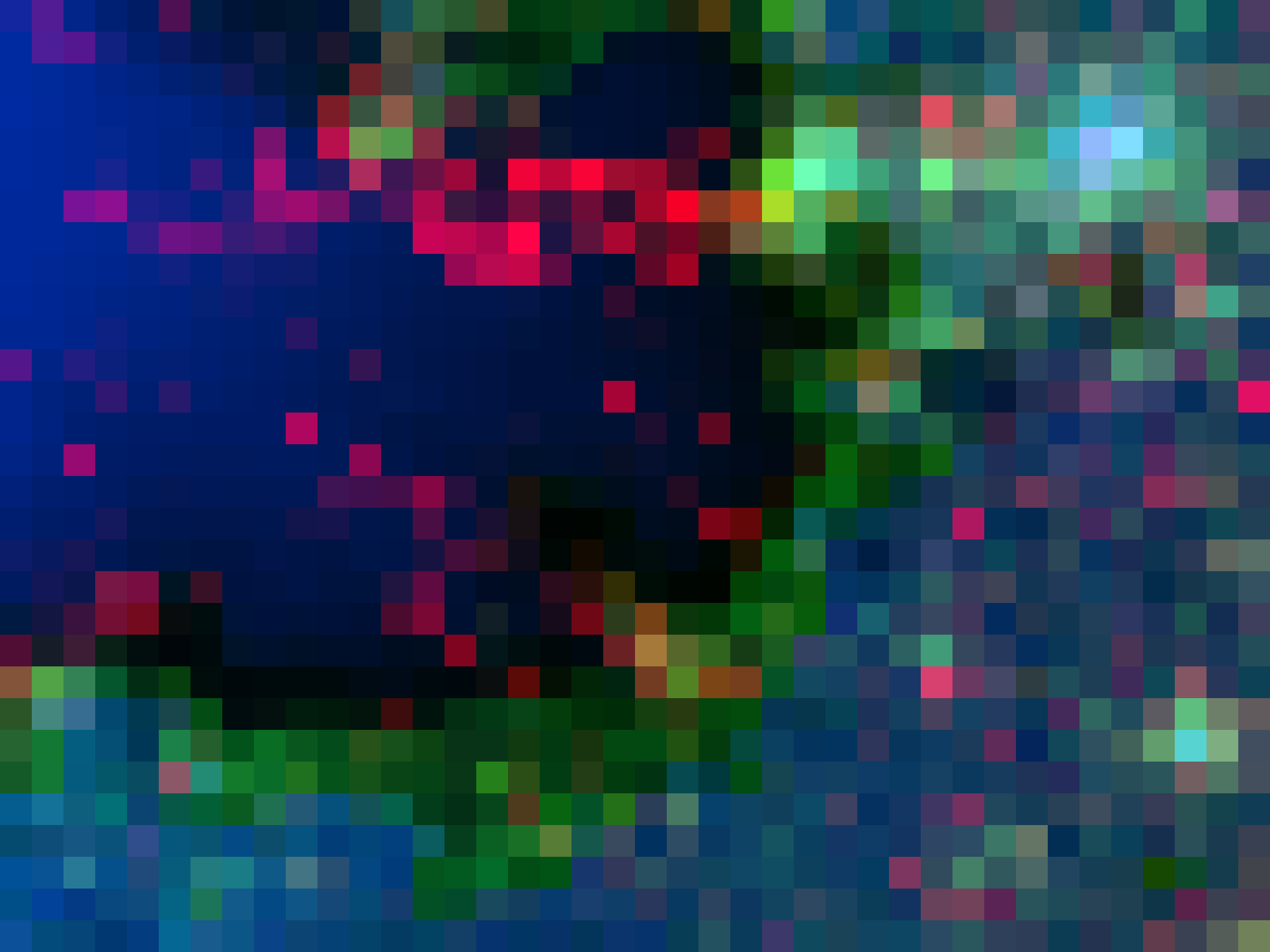}}
	\subcaptionbox{\label{Fig_segmentCIαSR_sub8}}
	{\includegraphics[width=0.32\linewidth]{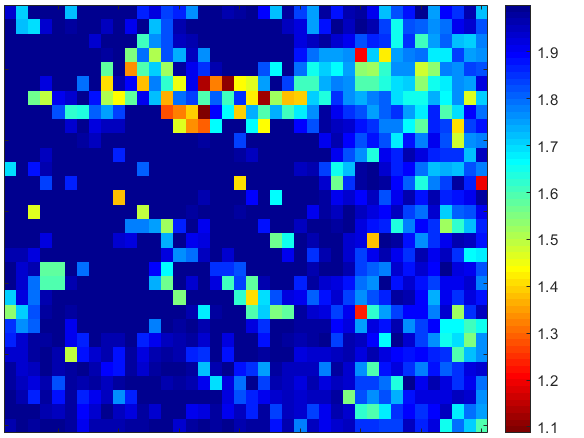}}
	\subcaptionbox{\label{Fig_segmentCIαSR_sub9}}
	{\includegraphics[width=0.32\linewidth]{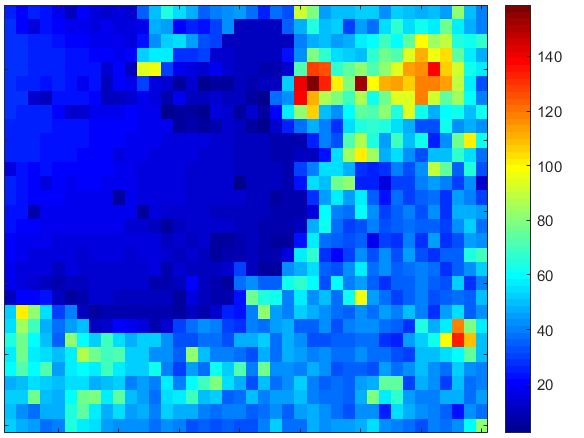}}
	\subcaptionbox{\label{Fig_segmentCIαSR_sub10}}
	{\includegraphics[width=0.32\linewidth]{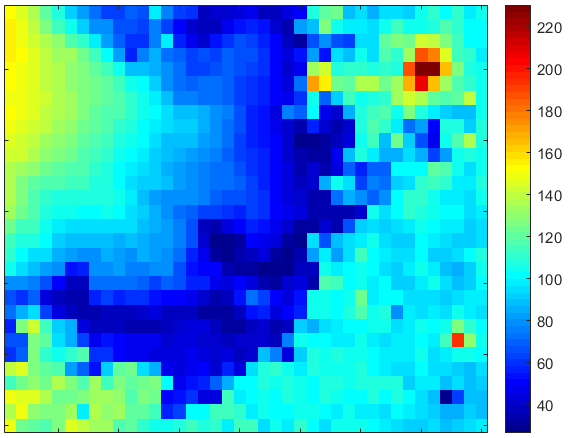}}
	\caption{Demonstration of CI$\alpha$SR's application in SAR feature extraction: (a) original SAR image of Shanghai and Eastern Jiangsu Province with grid indicating how patches are segmented; (c-e) heatmaps of estimated CI$\alpha$SR parameters $\alpha$, $\gamma$, and $\delta$, respectively; (b) psedo-RGB image generated from estimated parameters. (f-j) are similar counterparts for SAR images of portal city Tianjin.}
	\label{Fig_segmentCIαSR}
\end{figure}

In order to provide a practical interpretation of the CI$\alpha$SR model parameters, as well as excavating the model's potential applications in image classification and target detection, two large SAR images of Shanghai and Eastern Jiangsu Province in Figure \ref{Fig_segmentCIαSR_sub1} and portal city Tianjin in Figure \ref{Fig_segmentCIαSR_sub6} with different scenes combined are segmented into patches and the corresponding model parameters are estimated. Both images have an original dimension of 20000 by 1500 pixels that is equally divided into 40 by 30 square-shaped patches of 500 pixels wide.
The patches are processed by the proposed CI$\alpha$SR model to produce matrices of the three model parameters $\alpha$, $\gamma$, and $\delta$, which are displayed using heatmaps in Figure \ref{Fig_segmentCIαSR} (c-e) and (h-j). Pseudo-RGB images are subsequently generated by assigning the normalized value of the three parameters to red, green, and blue channel, respectively. Please note that both the colour bar for parameter $\alpha$ heatmaps and corresponding red channel intensity in pseudo-RGB images are inverted, since smaller value of $\alpha$ parameter contributes to data of heavier tail, and consequently a brighter pixel in the original images.

Comparison between the CI$\alpha$SR generated pseudo-RGB image and the original in Fig. \ref{Fig_segmentCIαSR} helps understand the meaning of CI$\alpha$SR parameters: characteristic exponent $\alpha$ represents the tail heaviness of data population, and red regions of small $\alpha$ in pseudo-RGB images indicates higher proportion of extremely large data values in contrast to the background, marking out major residential area in Figure \ref{Fig_segmentCIαSR_sub2} and voyaging ships in Figure \ref{Fig_segmentCIαSR_sub7}; scale parameter $\gamma$ represents the dispersion of data population, and green regions of large $\gamma$ in pseudo-RGB image usually indicate a mixture of different scenes which contribute to a histogram peak spread thin, marking out texture-complex patches that involve change of terrain such as lakeside in Figure \ref{Fig_segmentCIαSR_sub2} and coastline in Figure \ref{Fig_segmentCIαSR_sub7}; location parameter $\delta$ represents the average reflection intensity of the area, and blue regions of different shade indicate the intrinsic reflectance to SAR radio illumination, distinguishing land (high reflectance) from water (low reflectance). 
Sometimes all three channels are of high value to produce a white region, which corresponds to an extremely heterogeneous area with high base reflection and mixture of terrain, an example being the image segment in row 7 column 1 in Figure \ref{Fig_segmentCIαSR_sub2}, which is an extra crowded residential area with a river running pass it, this area coincides with the geographical landmark of Lujiazui, one of the most crowded districts in Shanghai. Likewise, it is possible to analyze image scene composition using the CI$\alpha$SR parameters for purpose of terrain classification and target detection.

\section{Conclusion}
A novel statistical model named CI$\alpha$SR for SAR images is proposed to cater scenes of various heterogeneity. The model describes the amplitude distribution of a complex isotropic $\alpha$-stable random variable, and is theoretically justified by a generalized Central Limit Theorem. Detailed experiments on both synthetic and real data have proved CI$\alpha$SR to be capable of faithfully representing the heavy tails of impulsive data population, which is particularly observed in SAR image data of urban area or sea loitered with ships. 
Alongside the model a quasi-analytical parameter estimator that combines a root-finding method and generalized method of Bessel moments (MoBM) is devised to acquire CI$\alpha$SR parameters at low computation cost. This generalization for the first time introduces Bessel function to the method of moments and may serve as a generic solution to other complicated distribution models. 
Physical significance of the CI$\alpha$SR models are explored by implementing the model on large-swath SAR images containing a mixture of different scenes. The three model parameter proves to be a good characterization of the heavy-tailedness, texture complexity, and average reflection of a SAR image, respectively. The proposed CI$\alpha$SR model demonstrates admirable potential in classification and target recognition of radar images and perhaps other types of image data.

Despite a dominant advantage in heterogeneous data, CI$\alpha$SR model falls short in modelling homogeneous data. One possible solution to compensate for this drawback is to further extend the model to include an adjustable skewness parameter $\beta$. However, a novel parameter estimator is required to support this more complicated model, which will be addressed by the authors in future work.

\bibliography{main}

\end{document}